\documentclass[aps,prl,superscriptaddress,twocolumn,tightenlines,nobibnotes]{revtex4-1}
\usepackage{graphicx,url,amssymb,amsmath,longtable,rotating,color,units,wasysym,subfigure,epsfig,hyperref,epstopdf}
\usepackage{acronym,cleveref}
\usepackage[dvipsnames]{xcolor}
\usepackage{pdfpages}

\crefformat{footnote}{#2\footnotemark[#1]#3}

\definecolor{darkgreen}{rgb}{0.0, 0.5, 0.0}

\newif\ifcomments
\commentstrue

\newcommand{\calP}{{\cal P}}
\newcommand{\calF}{{\cal F}}

\newcommand{\ohat}{\Theta} 
\newcommand{\phat}{\hat{\cal P}}
\newcommand{\ophat}{{\cal P}(\ohat)}

\newcommand{\Ogw}{\Omega_\text{gw}}
\newcommand{\rhogw}{\rho_\text{gw}}

\newcommand{\lmax}{l_\text{max}}
\newcommand{\falpha}{f_\alpha}
\newcommand{\fref}{f_\text{ref}}

\newcommand{\radP}{\calP_{\ohat_0}} 

\newcommand{\rundatesUTC}{{15:00 UTC, Sep 18,  2015 -- 16:00 UTC, Jan 12, 2016}}

\newcommand{\beststrain}{{\unit[$7\times10^{-24}$]{Hz$^{-1/2}$}}~}
\newcommand{\sensitiveband}{{\unit[100 -- 300]{Hz}}~}
\newcommand{\coincidence}{{\unit[49.67]{days}}~}
\newcommand{\sphUL}{~{$\Omega(f,\ohat) < \unit[(0.39 - 7.6)\times 10^{-8}]{\text{sr}^{-1}}(f/25 \text{ Hz})^{\alpha}$}~} 

\newcommand{\radUL}{~{$F_{\alpha,\Theta}(f) < \unit[(0.1 - 56)\times 10^{-8}]{\text{erg cm}^{-2}\text{ s}^{-1} \text{ Hz}^{-1}}(f/25 \text{ Hz})^{\alpha-1}$}~}

\newcommand{\radULnb}{{$h_0 < (6.7,\, 5.5, \text{ and } 7.0)\times10^{-25}$}~} 
\newcommand{\radULsco}{{$h_0 < \unit[6.7 \times10^{-25}]{\text{at 134 Hz}}$}~}
\newcommand{\radULsn}{{$h_0 < \unit[5.5 \times10^{-25}]{\text{at 172 Hz}}$}~}
\newcommand{\radULgc}{{$h_0 < \unit[7.0 \times10^{-25}]{\text{at 172 Hz}}$}~}
\def\nbfreq{130 -- 175 Hz} 

\newcommand{\fs}{f_\text{s}}
\newcommand{\fd}{f_\text{d}}
\newcommand{\fdet}{f_\text{d}}
\newcommand{\ve}{\vec{v}_\text{E}}
\newcommand{\Porb}{P_\text{orb}}
\newcommand{\Tasc}{T_\text{asc}}

\acrodef{LIGO}{Laser Interferometer Gravitational-wave Observatory}
\acrodef{aLIGO}{Advanced LIGO}
\acrodef{GW}{gravitational waves}
\acrodef{CGB}{cosmological gravitational-wave background}
\acrodef{SNR}{signal-to-noise ratio}

\newcommand{\dcc}{LIGO-P1600259}

\begin{document}

\title{Directional limits on persistent gravitational waves \\
from Advanced LIGO's first observing run}


%



\author{%
B.~P.~Abbott,$^{1}$  
R.~Abbott,$^{1}$  
T.~D.~Abbott,$^{2}$  
M.~R.~Abernathy,$^{3}$  
F.~Acernese,$^{4,5}$ 
K.~Ackley,$^{6}$  
C.~Adams,$^{7}$  
T.~Adams,$^{8}$ 
P.~Addesso,$^{9}$  
R.~X.~Adhikari,$^{1}$
% ...
\\
\medskip
(LIGO Scientific Collaboration and Virgo Collaboration) 
\\
\medskip
{{}$^{*}$Deceased, March 2016. }%
}\noaffiliation
\affiliation {LIGO, California Institute of Technology, Pasadena, CA 91125, USA }
\affiliation {Louisiana State University, Baton Rouge, LA 70803, USA }
\affiliation {American University, Washington, D.C. 20016, USA }
\affiliation {Universit\`a di Salerno, Fisciano, I-84084 Salerno, Italy }
\affiliation {INFN, Sezione di Napoli, Complesso Universitario di Monte S.Angelo, I-80126 Napoli, Italy }
\affiliation {University of Florida, Gainesville, FL 32611, USA }
\affiliation {LIGO Livingston Observatory, Livingston, LA 70754, USA }
\affiliation {Laboratoire d'Annecy-le-Vieux de Physique des Particules (LAPP), Universit\'e Savoie Mont Blanc, CNRS/IN2P3, F-74941 Annecy-le-Vieux, France }
\affiliation {University of Sannio at Benevento, I-82100 Benevento, Italy and INFN, Sezione di Napoli, I-80100 Napoli, Italy }
\affiliation {Albert-Einstein-Institut, Max-Planck-Institut f\"ur Gravi\-ta\-tions\-physik, D-30167 Hannover, Germany }
\affiliation {Nikhef, Science Park, 1098 XG Amsterdam, The Netherlands }
\affiliation {LIGO, Massachusetts Institute of Technology, Cambridge, MA 02139, USA }
\affiliation {Instituto Nacional de Pesquisas Espaciais, 12227-010 S\~{a}o Jos\'{e} dos Campos, S\~{a}o Paulo, Brazil }
\affiliation {INFN, Gran Sasso Science Institute, I-67100 L'Aquila, Italy }
\affiliation {INFN, Sezione di Roma Tor Vergata, I-00133 Roma, Italy }
\affiliation {Inter-University Centre for Astronomy and Astrophysics, Pune 411007, India }
\affiliation {International Centre for Theoretical Sciences, Tata Institute of Fundamental Research, Bengaluru 560089, India }
\affiliation {University of Wisconsin-Milwaukee, Milwaukee, WI 53201, USA }
\affiliation {Leibniz Universit\"at Hannover, D-30167 Hannover, Germany }
\affiliation {Universit\`a di Pisa, I-56127 Pisa, Italy }
\affiliation {INFN, Sezione di Pisa, I-56127 Pisa, Italy }
\affiliation {Australian National University, Canberra, Australian Capital Territory 0200, Australia }
\affiliation {California State University Fullerton, Fullerton, CA 92831, USA }
\affiliation {LAL, Univ. Paris-Sud, CNRS/IN2P3, Universit\'e Paris-Saclay, F-91898 Orsay, France }
\affiliation {Chennai Mathematical Institute, Chennai 603103, India }
\affiliation {Universit\`a di Roma Tor Vergata, I-00133 Roma, Italy }
\affiliation {Universit\"at Hamburg, D-22761 Hamburg, Germany }
\affiliation {INFN, Sezione di Roma, I-00185 Roma, Italy }
\affiliation {Albert-Einstein-Institut, Max-Planck-Institut f\"ur Gravitations\-physik, D-14476 Potsdam-Golm, Germany }
\affiliation {APC, AstroParticule et Cosmologie, Universit\'e Paris Diderot, CNRS/IN2P3, CEA/Irfu, Observatoire de Paris, Sorbonne Paris Cit\'e, F-75205 Paris Cedex 13, France }
\affiliation {West Virginia University, Morgantown, WV 26506, USA }
\affiliation {Universit\`a di Perugia, I-06123 Perugia, Italy }
\affiliation {INFN, Sezione di Perugia, I-06123 Perugia, Italy }
\affiliation {European Gravitational Observatory (EGO), I-56021 Cascina, Pisa, Italy }
\affiliation {Syracuse University, Syracuse, NY 13244, USA }
\affiliation {SUPA, University of Glasgow, Glasgow G12 8QQ, United Kingdom }
\affiliation {LIGO Hanford Observatory, Richland, WA 99352, USA }
\affiliation {Wigner RCP, RMKI, H-1121 Budapest, Konkoly Thege Mikl\'os \'ut 29-33, Hungary }
\affiliation {Columbia University, New York, NY 10027, USA }
\affiliation {Stanford University, Stanford, CA 94305, USA }
\affiliation {Universit\`a di Padova, Dipartimento di Fisica e Astronomia, I-35131 Padova, Italy }
\affiliation {INFN, Sezione di Padova, I-35131 Padova, Italy }
\affiliation {Nicolaus Copernicus Astronomical Center, Polish Academy of Sciences, 00-716, Warsaw, Poland }
\affiliation {Center for Relativistic Astrophysics and School of Physics, Georgia Institute of Technology, Atlanta, GA 30332, USA }
\affiliation {University of Birmingham, Birmingham B15 2TT, United Kingdom }
\affiliation {Universit\`a degli Studi di Genova, I-16146 Genova, Italy }
\affiliation {INFN, Sezione di Genova, I-16146 Genova, Italy }
\affiliation {RRCAT, Indore MP 452013, India }
\affiliation {Faculty of Physics, Lomonosov Moscow State University, Moscow 119991, Russia }
\affiliation {SUPA, University of the West of Scotland, Paisley PA1 2BE, United Kingdom }
\affiliation {Caltech CaRT, Pasadena, CA 91125, USA }
\affiliation {University of Western Australia, Crawley, Western Australia 6009, Australia }
\affiliation {Department of Astrophysics/IMAPP, Radboud University Nijmegen, P.O. Box 9010, 6500 GL Nijmegen, The Netherlands }
\affiliation {Artemis, Universit\'e C\^ote d'Azur, CNRS, Observatoire C\^ote d'Azur, CS 34229, F-06304 Nice Cedex 4, France }
\affiliation {Institut de Physique de Rennes, CNRS, Universit\'e de Rennes 1, F-35042 Rennes, France }
\affiliation {Washington State University, Pullman, WA 99164, USA }
\affiliation {Universit\`a degli Studi di Urbino 'Carlo Bo', I-61029 Urbino, Italy }
\affiliation {INFN, Sezione di Firenze, I-50019 Sesto Fiorentino, Firenze, Italy }
\affiliation {University of Oregon, Eugene, OR 97403, USA }
\affiliation {Laboratoire Kastler Brossel, UPMC-Sorbonne Universit\'es, CNRS, ENS-PSL Research University, Coll\`ege de France, F-75005 Paris, France }
\affiliation {Carleton College, Northfield, MN 55057, USA }
\affiliation {Astronomical Observatory Warsaw University, 00-478 Warsaw, Poland }
\affiliation {VU University Amsterdam, 1081 HV Amsterdam, The Netherlands }
\affiliation {University of Maryland, College Park, MD 20742, USA }
\affiliation {Laboratoire des Mat\'eriaux Avanc\'es (LMA), CNRS/IN2P3, F-69622 Villeurbanne, France }
\affiliation {Universit\'e Claude Bernard Lyon 1, F-69622 Villeurbanne, France }
\affiliation {Universit\`a di Napoli 'Federico II', Complesso Universitario di Monte S.Angelo, I-80126 Napoli, Italy }
\affiliation {NASA/Goddard Space Flight Center, Greenbelt, MD 20771, USA }
\affiliation {RESCEU, University of Tokyo, Tokyo, 113-0033, Japan. }
\affiliation {University of Adelaide, Adelaide, South Australia 5005, Australia }
\affiliation {Tsinghua University, Beijing 100084, China }
\affiliation {Texas Tech University, Lubbock, TX 79409, USA }
\affiliation {The University of Mississippi, University, MS 38677, USA }
\affiliation {The Pennsylvania State University, University Park, PA 16802, USA }
\affiliation {National Tsing Hua University, Hsinchu City, 30013 Taiwan, Republic of China }
\affiliation {Charles Sturt University, Wagga Wagga, New South Wales 2678, Australia }
\affiliation {University of Chicago, Chicago, IL 60637, USA }
\affiliation {Kenyon College, Gambier, OH 43022, USA }
\affiliation {Korea Institute of Science and Technology Information, Daejeon 305-806, Korea }
\affiliation {University of Cambridge, Cambridge CB2 1TN, United Kingdom }
\affiliation {Universit\`a di Roma 'La Sapienza', I-00185 Roma, Italy }
\affiliation {University of Brussels, Brussels 1050, Belgium }
\affiliation {Sonoma State University, Rohnert Park, CA 94928, USA }
\affiliation {Montana State University, Bozeman, MT 59717, USA }
\affiliation {Center for Interdisciplinary Exploration \& Research in Astrophysics (CIERA), Northwestern University, Evanston, IL 60208, USA }
\affiliation {Universitat de les Illes Balears, IAC3---IEEC, E-07122 Palma de Mallorca, Spain }
\affiliation {The University of Texas Rio Grande Valley, Brownsville, TX 78520, USA }
\affiliation {Bellevue College, Bellevue, WA 98007, USA }
\affiliation {Institute for Plasma Research, Bhat, Gandhinagar 382428, India }
\affiliation {The University of Sheffield, Sheffield S10 2TN, United Kingdom }
\affiliation {California State University, Los Angeles, 5154 State University Dr, Los Angeles, CA 90032, USA }
\affiliation {Universit\`a di Trento, Dipartimento di Fisica, I-38123 Povo, Trento, Italy }
\affiliation {INFN, Trento Institute for Fundamental Physics and Applications, I-38123 Povo, Trento, Italy }
\affiliation {Cardiff University, Cardiff CF24 3AA, United Kingdom }
\affiliation {Montclair State University, Montclair, NJ 07043, USA }
\affiliation {National Astronomical Observatory of Japan, 2-21-1 Osawa, Mitaka, Tokyo 181-8588, Japan }
\affiliation {Canadian Institute for Theoretical Astrophysics, University of Toronto, Toronto, Ontario M5S 3H8, Canada }
\affiliation {MTA E\"otv\"os University, ``Lendulet'' Astrophysics Research Group, Budapest 1117, Hungary }
\affiliation {School of Mathematics, University of Edinburgh, Edinburgh EH9 3FD, United Kingdom }
\affiliation {University and Institute of Advanced Research, Gandhinagar, Gujarat 382007, India }
\affiliation {IISER-TVM, CET Campus, Trivandrum Kerala 695016, India }
\affiliation {University of Szeged, D\'om t\'er 9, Szeged 6720, Hungary }
\affiliation {Embry-Riddle Aeronautical University, Prescott, AZ 86301, USA }
\affiliation {Tata Institute of Fundamental Research, Mumbai 400005, India }
\affiliation {INAF, Osservatorio Astronomico di Capodimonte, I-80131, Napoli, Italy }
\affiliation {University of Michigan, Ann Arbor, MI 48109, USA }
\affiliation {Rochester Institute of Technology, Rochester, NY 14623, USA }
\affiliation {NCSA, University of Illinois at Urbana-Champaign, Urbana, IL 61801, USA }
\affiliation {University of Bia{\l }ystok, 15-424 Bia{\l }ystok, Poland }
\affiliation {SUPA, University of Strathclyde, Glasgow G1 1XQ, United Kingdom }
\affiliation {University of Southampton, Southampton SO17 1BJ, United Kingdom }
\affiliation {University of Washington Bothell, 18115 Campus Way NE, Bothell, WA 98011, USA }
\affiliation {Institute of Applied Physics, Nizhny Novgorod, 603950, Russia }
\affiliation {Seoul National University, Seoul 151-742, Korea }
\affiliation {Inje University Gimhae, 621-749 South Gyeongsang, Korea }
\affiliation {National Institute for Mathematical Sciences, Daejeon 305-390, Korea }
\affiliation {Pusan National University, Busan 609-735, Korea }
\affiliation {NCBJ, 05-400 \'Swierk-Otwock, Poland }
\affiliation {Institute of Mathematics, Polish Academy of Sciences, 00656 Warsaw, Poland }
\affiliation {The School of Physics \& Astronomy, Monash University, Clayton 3800, Victoria, Australia }
\affiliation {Hanyang University, Seoul 133-791, Korea }
\affiliation {The Chinese University of Hong Kong, Shatin, NT, Hong Kong }
\affiliation {University of Alabama in Huntsville, Huntsville, AL 35899, USA }
\affiliation {ESPCI, CNRS, F-75005 Paris, France }
\affiliation {University of Minnesota, Minneapolis, MN 55455, USA }
\affiliation {Universit\`a di Camerino, Dipartimento di Fisica, I-62032 Camerino, Italy }
\affiliation {Southern University and A\&M College, Baton Rouge, LA 70813, USA }
\affiliation {The University of Melbourne, Parkville, Victoria 3010, Australia }
\affiliation {College of William and Mary, Williamsburg, VA 23187, USA }
\affiliation {Instituto de F\'\i sica Te\'orica, University Estadual Paulista/ICTP South American Institute for Fundamental Research, S\~ao Paulo SP 01140-070, Brazil }
\affiliation {Whitman College, 345 Boyer Avenue, Walla Walla, WA 99362 USA }
\affiliation {Universit\'e de Lyon, F-69361 Lyon, France }
\affiliation {Hobart and William Smith Colleges, Geneva, NY 14456, USA }
\affiliation {Janusz Gil Institute of Astronomy, University of Zielona G\'ora, 65-265 Zielona G\'ora, Poland }
\affiliation {King's College London, University of London, London WC2R 2LS, United Kingdom }
\affiliation {IISER-Kolkata, Mohanpur, West Bengal 741252, India }
\affiliation {Indian Institute of Technology, Gandhinagar Ahmedabad Gujarat 382424, India }
\affiliation {Andrews University, Berrien Springs, MI 49104, USA }
\affiliation {Universit\`a di Siena, I-53100 Siena, Italy }
\affiliation {Trinity University, San Antonio, TX 78212, USA }
\affiliation {University of Washington, Seattle, WA 98195, USA }
\affiliation {Abilene Christian University, Abilene, TX 79699, USA }



\noaffiliation


\begin{abstract}
We employ gravitational-wave radiometry to map the gravitational waves stochastic background expected from a variety of contributing mechanisms and test the assumption of isotropy using data from Advanced LIGO's first observing run. 
We also search for persistent gravitational waves from point sources with only minimal assumptions over the 20 - 1726 Hz frequency band.
Finding no evidence of gravitational waves from either point sources or a stochastic background, we set limits at {90\%} confidence.
For broadband point sources, we report upper limits on the gravitational wave energy flux per unit frequency in the range \radUL depending on the sky location $\ohat$ and the spectral power index $\alpha$.  
For extended sources, we report upper limits on the fractional gravitational wave energy density required to close the Universe of \sphUL depending on $\ohat$ and $\alpha$.
Directed searches for narrowband gravitational waves from astrophysically interesting objects (Scorpius X-1, Supernova 1987~A, and the Galactic Center) yield median frequency-dependent limits on strain amplitude of \radULnb respectively, at the most sensitive detector frequencies between \nbfreq.
This represents a mean improvement of a factor of 2 across the band compared to previous searches of this kind for these sky locations, considering the different quantities of strain constrained in each case.
\end{abstract}

\maketitle


\acrodef{BNS}{Binary Neutron Star}
\acrodef{BBH}{Binary Black Hole}
\acrodef{SNR}{signal-to-noise ratio}
\acrodef{GW}{gravitational wave}

{\em Introduction.}---A stochastic gravitational-wave background (SGWB) is expected from a variety of mechanisms \cite{Maggiore,  Allen_1997, Mandic_Buonanno_2006, ZhuEA_2011_rm, SiemensEA_2007}.
Given the recent observations of binary black hole mergers GW150914 and GW151226 \cite{gw150914, gw151226}, 
we expect the SGWB to be nearly isotropic \cite{meacher} and dominated \cite{CallisterEA_2016} by compact binary coalescences \cite{tania_cbc2,WuEA_2012,ZhuEA_2013}.
The LIGO and Virgo Collaborations have pursued the search for an isotropic stochastic background from LIGO's first observational run~\cite{o1iso}.
Here, we adopt an “eyes-wide-open” philosophy and relax the assumption of isotropy in order to allow for the greater range of possible signals. 
We search for an anisotropic background, which could indicate a richer, more interesting cosmology than current models.
We present the results of a generalized search for a stochastic signal with an arbitrary angular distribution mapped over all directions in the sky. 

Our search has three components. 
First, we utilize a broadband radiometer analysis~\cite{radio_method,radiometer}, optimized for detecting a small number of resolvable point sources.
This method is not applicable to extended sources.
Second, we employ a spherical harmonic decomposition ~\cite{sph_methods, sph_results}, which can be employed for point sources but is better suited to extended sources.
Last, we carry out a narrowband radiometer search directed at the sky position of three astrophysically interesting objects: Scorpius X-1 (Sco X-1) \cite{scox1_s5, scox1_mdc}, Supernova 1987~A (SN 1987A) \cite{ChungEA_sn1987a,SunEA_1987a}, and the Galactic Center (GC) \cite{gc_2013}.

These three search methods are capable of detecting a wide range of possible signals with only minimal assumptions about the signal morphology.
We find no evidence of persistent gravitational waves, and set limits on broadband emission of gravitational waves as a function of sky position. 
We also set narrowband limits as a function of frequency for the three selected sky positions.

{\em Data.}---We analyze data from Advanced LIGO's $\unit[4]{km}$ detectors in Hanford, WA (H1) and Livingston, LA (L1) during the first observing run (O1), from \rundatesUTC. During O1, the detectors reached an instantaneous strain sensitivity of \beststrain in the most sensitive region between \sensitiveband, and collected \coincidence of coincident H1L1 data. The O1 observing run saw the first direct detection of gravitational waves and the first direct observation of merging black holes \cite{gw150914, gw151226}. 

For our analysis, the time-series data are down-sampled to 4096 Hz from 16 kHz, and divided into $\unit[192]{s}$, 50\% overlapping, Hann-windowed segments, which are high-pass filtered with a 16th order Butterworth digital filter with knee frequency of 11 Hz (following \cite{s4iso,o1iso}).
We apply data quality cuts in the time domain in order to remove segments associated with instrumental artifacts and hardware injections used for signal validation~\cite{gw150914_detchar, o1hwinj}.
We also exclude segments containing known gravitational-wave signals. 
Finally, we apply a standard non-stationarity cut (see, e.g.,~\cite{stoch-S5}), to eliminate segments that do not behave as Gaussian noise.
These cuts remove 35\% of the data. With all vetoes applied, the total live time for 192 s segments is 29.85 days.

The data segments are Fourier transformed and coarse-grained to produce power spectra with a resolution of $\unit[1/32]{Hz}$.
This is a finer frequency resolution than the $\unit[1/4]{Hz}$ used in previous LIGO/Virgo stochastic searches~\cite{radiometer, sph_results} in order to remove many finely spaced instrumental lines occurring at low frequencies.
Frequency bins associated with known instrumental artifacts including suspension violin modes \cite{aLIGOsuspension2012}, calibration lines, electronic lines, correlated combs, and signal injections of persistent, monochromatic, gravitational waves are not included in the analysis.
These frequency domain cuts remove 21\% of the observing band.
For a detailed description of data quality studies performed for this analysis, see the supplement \cite{technical_supplement} of \cite{o1iso}.

The broadband searches include frequencies from 20 -- 500 Hz which more than cover the regions of $99\%$ sensitivity for each of the spectral bands (see Table 1 of \cite{o1iso}).
The narrowband analysis covers the full 20 -- 1726 Hz band.

{\em Method.}---
The main goal of a stochastic search is to estimate the fractional contribution of the energy density in gravitational waves $\Ogw$ to the total energy density needed to close the Universe $\rho_c$. 
This is defined by
\begin{equation}
  \Ogw(f) = \frac{f}{\rho_c} \frac{d\rhogw}{d f} 
\end{equation}
where $f$ is frequency and $d\rho_\text{gw}$ represents the energy density between $f$ and $f+ df$ \cite{allen-romano}. 
For a stationary and unpolarized signal, $\rho_\text{gw}$ can be factored into an angular power $\ophat$ and a spectral shape $H(f)$ \cite{allen-ottewill}, such that
\begin{equation}
  \Ogw(f) = \frac{2\pi^2}{3H_0^2} f^3 H(f) \int d\ohat\>\ophat, \label{eq:Ogw}
\end{equation}
with Hubble constant $H_0=\unit[68]{km\,s^{-1}\,Mpc^{-1}}$ from~\cite{Planck_2014}.

The angular power $\ophat$ represents the gravitational wave power at each point in the sky. To express this in terms of the fractional energy density, we define the energy density spectrum as a function of sky position 
\begin{equation}
\Omega(f,\ohat) = \frac{2\pi^2}{3H_0^2}f^3{H(f)} \ophat.
\end{equation}
We define a similar quantity for the energy flux, where
\begin{equation}
\calF(f,\ohat) = \frac{c^3 \pi }{4G}f^2{H(f)}\ophat
\end{equation}
has units of $\text{erg cm}^{-2}\text{ s}^{-1}\text{ Hz}^{-1}{\text{ sr}^{-1}}$~\cite{radiometer, sph_methods}, $c$ is the speed of light and $G$ is Newton's gravitational constant.

{\em Point sources versus extended sources.}---We employ two different methods to estimate $\ophat$ based on the cross-correlation of data streams from a pair of detectors \cite{allen-romano, sph_results}. 
The radiometer method~\cite{radiometer,radio_method} assumes that the cross-correlation signal is dominated by a small number of resolvable point sources.
The point source power is given by $\radP$ and the angular power spectrum is then 
\begin{equation}\label{eq:rad_basis}
  \ophat \equiv \radP \delta^2(\ohat, \ohat_0) .
\end{equation}

Although the radiometer method provides the optimal method for detecting resolvable point sources, it is not well-suited for describing diffuse or extended sources, which may have an arbitrary angular distribution. Hence, we also implement a complementary spherical harmonic decomposition (SHD) algorithm, in which the sky map is decomposed into components $Y_{lm}(\ohat)$ with coefficients ${\cal P}_{lm}$~\cite{sph_methods}:
\begin{equation}
  \ophat \equiv \sum_{lm} {\cal P}_{lm} Y_{lm}(\ohat) .
\end{equation}
Here, the sum over $l$ runs from $0$ to $l_\text{max}$ and $-l \leq m \leq l$.
We discuss the choice of $l_\text{max}$ below.
While the SHD algorithm has comparably worse sensitivity to point sources than the radiometer algorithm, it accounts for the detector response, producing more accurate sky maps.
 
{\em Spectral models.}---In both the radiometer algorithm and the spherical harmonic decomposition algorithm, we must choose a spectral shape $H(f)$.
We model the spectral dependence of $\Ogw(f)$ as a power law:
\begin{equation} \label{eq:Hf} 
H(f) = {\left( \frac{f}{\fref} \right)}^{\alpha-3},
\end{equation}
where $\fref$ is an arbitrary reference frequency and $\alpha$ is the spectral index (see also \cite{o1iso}).
The spectral model will also affect the angular power spectrum, so $\ophat$ is implicitly a function of $\alpha$.

We can rewrite the energy density map $\Omega(f,\ohat)$ to emphasize the spectral properties, such that
\begin{equation}
\Omega(f,\ohat) = \Omega_\alpha(\ohat) {\left(\frac{f}{\fref}\right)}^{\alpha},
\end{equation}
where 
\begin{equation} \label{eq:Omega_alpha}
  \Omega_\alpha(\ohat) = \frac{2\pi^2}{3H_0^2}\fref^3\calP(\ohat) 
\end{equation}
has units of fractional energy density per steradian $\Ogw \text{ sr}^{-1}$.
The spherical harmonic analysis presents skymaps of $\Omega_\alpha(\ohat)$. Note that when $\ophat = {\cal P}_{00}$ (the monopole moment), we recover a measurement for the energy density of the isotropic gravitational wave background.
Similarly, the gravitational wave energy flux can be expressed as
\begin{equation}
{\calF}(f,\ohat) = \calF_\alpha(\ohat) {\left(\frac{f}{\fref}\right)}^{\alpha-1},
\end{equation} 
where
\begin{equation}\label{eq:flux_alpha}
\calF_\alpha(\ohat) = \frac{c^3\pi}{4G}\fref^2\calP(\ohat). 
\end{equation}
In the radiometer case we calculate the flux in each direction 
\begin{equation}\label{eq:flux_radiometer}
        F_{\alpha,\Theta_0} = \frac{c^3\pi}{4G}\fref^2 \radP,
\end{equation}
which is obtained by integrating Equation~\ref{eq:flux_alpha} over the sphere for the point-source signal model described in Equation~\ref{eq:rad_basis}.
This quantity has units of $\text{erg cm}^{-2}\text{ s}^{-1}\text{ Hz}^{-1}.$
Following \cite{gw150914_stoch, o1iso}, we choose $\fref = 25$ Hz, corresponding to the most sensitive frequency in the spectral band for a stochastic search with the Advanced LIGO network at design sensitivity.
 
We consider three spectral indices: $\alpha=0$, corresponding to a flat energy density spectrum (expected from models of a cosmological background), $\alpha=2/3$, corresponding to the expected shape from a population of compact binary coalescences, and $\alpha=3$, corresponding to a flat strain power spectral density spectrum \cite{sph_results,  gw150914_stoch}. 
The different spectral models are summarized in Table~\ref{tab:alphas}. 

\begin{table*}
  \begin{tabular}{ccc | cc | ccc||cc|cc}
   \multicolumn{8}{}{}& \multicolumn{4}{c}{\bf All-sky (broadband) Results} \\
  \multicolumn{8}{}{}& \multicolumn{2}{c}{Max SNR  (\% $p$-value)} & \multicolumn{2}{c}{Upper limit range } \\
    \hline
    &$\alpha$&  & $\Ogw$ & $H(f)$  & $\falpha$ (Hz) & $\theta \,(\deg)$ & $\lmax$ & BBR &  SHD   & BBR ($\times 10^{-8}$) & SHD ($\times 10^{-8}$)\\
    \hline 
    &0&			& constant 			&  $\propto f^{-3}$       & 52.50   & 55 & 3   & 3.32  (7) & 2.69 (18) & 10 -- 56   & 2.5 -- 7.6  \\
    &$2/3$& 	& $\propto f^{2/3}$	&  $\propto f^{-7/3}$     & 65.75   & 44 & 4   & 3.31 (12) & 3.06 (11) & 5.1 -- 33  & 2.0 -- 5.9  \\
    &3&			& $\propto f^{3}$	&  constant       		  & 256.50  & 11 & 16  & 3.43 (47) & 3.86 (11) & 0.1 -- 0.9 & 0.4 -- 2.8  \\
    \hline
  \end{tabular}
  \caption{Values of the power-law index $\alpha$ investigated in this analysis, the shape of the energy density and strain power spectrum. The characteristic frequency $f_\alpha$, angular resolution $\theta$ (Eq. \ref{eq:theta}), and corresponding harmonic order $\lmax$ (Eq. \ref{eq:lmax}) for each $\alpha$ are also shown. The right hand section of the table shows the maximum SNR, associated significance ($p$-value) and best upper limit values from the broadband radiometer (BBR) and the spherical harmonic decomposition (SHD). The BBR sets upper limits on energy flux [${\text{erg cm}^{-2}\text{ s}^{-1} \text{ Hz}^{-1}}(f/25 \text{ Hz})^{\alpha-1}$] 
  while the SHD sets upper limits on the normalized energy density [${\text{sr}^{-1}}(f/25 \text{ Hz})^{\alpha}$] 
  of the SGWB.
  } \label{tab:alphas}
\end{table*}

{\em Cross Correlation.}--- A stochastic background would induce low-level correlation between the two LIGO detectors.
Although the signal is expected to be buried in the detector noise, the cross-correlation \ac{SNR} grows with the square root of integration time~\cite{allen-romano}.
The cross correlation between two detectors, with (one-sided strain) power spectral density $P_i(f,t)$ for detector $i$, is encoded in what is known as ``the dirty map''~\cite{sph_methods}:
\begin{equation} \label{eq:Xnu}
    X_\nu = \sum_{ft} \gamma^*_\nu(f,t) \frac{H(f)}{P_1(f,t)P_2(f,t)} C(f,t) . 
\end{equation} 
Here, $\nu$ is an index, which can refer to either individual points on the sky (the pixel basis) or different $lm$ indices (the spherical harmonic basis).
The variable $C(f,t)$ is the cross-power spectral density measured between the two LIGO detectors at some segment time $t$.
The sum runs over all segment times and all frequency bins.
The variable $\gamma_\nu(f,t)$ is a generalization of the overlap reduction function, which is a function of the separation and relative orientation between the detectors, and characterizes the frequency response of the detector pair~\cite{nelson}; see~\cite{sph_methods} for an exact definition.

We can think of $X_\nu$ as a sky map representation of the raw cross-correlation measurement before deconvolving the detector response.
The associated uncertainty is encoded in the Fisher matrix:
\begin{equation} \label{eq:Fisher}
  \Gamma_{\mu\nu} = 
  \sum_{ft} \gamma^*_\mu(f,t) \frac{H^2(f)}{P_1(f,t)P_2(f,t)} \gamma_\nu(f,t),
\end{equation} 
where ${}^*$ denotes complex conjugation. 
  
Once $X_\nu$ and $\Gamma_{\mu\nu}$ are calculated, we have the ingredients to calculate both the radiometer map and the SHD map.
However, the inversion of $\Gamma_{\mu\nu}$ is required to calculate the maximum likelihood estimators of GW power $\phat_\mu = \Gamma_{\mu\nu}^{-1}X_{\nu}$ \cite{sph_methods}.
For the radiometer, the correlations between neighbouring pixels can be ignored.
The radiometer map is given by 
\begin{equation}
  \begin{split}
    \phat_{\ohat} = & (\Gamma_{\ohat\ohat})^{-1} X_{\ohat}\\
    \sigma_{\ohat}^\text{rad} = & (\Gamma_{\ohat\ohat})^{-1/2},
  \end{split}
\end{equation}
where the standard deviation $\sigma_{\ohat}^\text{rad}$ is the uncertainty associated with the point source amplitude estimator $\phat_{\ohat}$,
and $\Gamma_{\ohat\ohat}$ is a diagonal entry of the Fisher matrix for a pointlike signal.
For the SHD analysis, the full Fisher matrix $\Gamma_{\mu\nu}$ must be taken into account, which includes singular eigenvalues associated with modes to which the detector pair is insensitive.
The inversion of $\Gamma_{\mu\nu}$ is simplified by a singular value decomposition regularization. 
In this decomposition, modes associated with the smallest eigenvalues contribute the least sensitivity to the detector network. 
Removing a fraction of the lowest eigenmodes ``regularizes" $\Gamma_{\mu\nu}$ without significantly affecting the sensitivity (see \cite{sph_methods}). 
The estimator for the SHD and corresponding standard deviation are given by
\begin{equation}\label{eq:Plm}
  \begin{split}
        \hat{\cal P}_{lm} = & \sum_{l'm'} (\Gamma_R^{-1})_{lm,l'm'} X_{l'm'} \\
        \sigma_{lm}^\text{SHD} = & \left[(\Gamma_R^{-1})_{lm,lm}\right]^{1/2} , 
  \end{split}
\end{equation}
where $\Gamma_R$ is the \emph{regularized} Fisher matrix.  
We remove $1/3$ of the lowest eigenvalues following~\cite{sph_methods, sph_results}.

{\em Angular scale.}---In order to carry out the calculation in Eq.~\ref{eq:Plm}, 
we must determine a suitable angular scale, which will depend on the angular resolution of the detector network and vary with spectral index $\alpha$. 
The diffraction-limited spot size on the sky $\theta$ (in radians) is given by
\begin{equation} \label{eq:theta}
\theta = \frac{c}{2df} \approx \frac{50 \text{ Hz}}{f_\alpha} ,
\end{equation}
where $d=3000$ km is the separation of the LIGO detectors. 
The frequency $f_\alpha$ corresponds to the most sensitive frequency in the detector band for a power law with spectral index $\alpha$ given the detector noise power spectra \cite{radiometer}.
In order to determine $f_\alpha$ we find the frequency at which a power-law with index $\alpha$ is tangent to the single-detector ``power-law integrated curve''~\cite{locus}. 
The angular resolution scale is set by the maximum spherical harmonic order $\lmax$, which we can express as a function of $\alpha$ since
\begin{equation} \label{eq:lmax}
\lmax = \frac{\pi}{\theta} \approx \frac{\pi f_\alpha}{50 \text{Hz}}.
\end{equation}
The values of $f_\alpha$, $\theta$, and $\lmax$ for three different values of $\alpha$ are shown in Table~\ref{tab:alphas}.
As the spectral index increases, so does $f_\alpha$, decreasing the angular resolution limit, thus increasing $\lmax$.

{\em Angular power spectra.}---For the SHD map, we calculate the angular power spectra $C_l$, which describe the angular scale of structure in the clean map,
using an unbiased estimator ~\cite{sph_results,sph_methods}
\begin{equation}
 \hat C_l \equiv  \frac{1}{2l+1} \sum_m \left[|\hat{\cal P}_{lm}|^2 - (\Gamma_R^{-1})_{lm,lm } \right] .
\end{equation}
%

{\em Narrowband radiometer.}---The radiometer algorithm can be applied to the detection of persistent gravitational waves from narrowband point sources associated with a given sky position \cite{radiometer, sph_results}. 
We ``point'' the radiometer in the direction of three interesting sky locations: Sco X-1, the Galactic Center, and the remnant of supernova SN 1987A.

Scorpius X-1 (Sco X-1) is a low-mass X-ray binary believed to host a neutron star that is potentially spun up through accretion, in which gravitational wave emission may provide a balancing spin-down torque \cite{Bildsten_1998, scox1_s2, scox1_s5, scox1_mdc}.
The frequency of the gravitational wave signal is expected to spread due to the orbital motion of the neutron star. 
At frequencies below $\sim$\unit[930]{Hz} this Doppler line broadening effect is less than \unit[1/4]{Hz}, the frequency bin width selected in past analyses~\cite{radiometer,sph_results}. 
At higher frequencies, the signal is certain to span multiple bins. 
We therefore combine multiple \unit[1/32]{Hz} frequency bins to form optimally sized \emph{combined bins} at each frequency, accounting for the expected signal broadening due to the combination of the motion of the Earth around the Sun, the binary orbital motion, and any other intrinsic modulation. For more detail on the method of combining bins, see the technical supplement to this paper \cite{supplement_nb}.

The possibility of a young neutron star in SN 1987A \cite{ChungEA_sn1987a, SunEA_1987a}
and the likelihood of many unknown, isolated neutron stars in the Galactic Center region \cite{gc_2013}
indicate potentially interesting candidates for persistent gravitational wave emission.
We combine bins to include the signal spread due to Earth's modulation.
For SN 1987A, we choose a combined bin size of 0.09 Hz. We would be sensitive to spin modulations up to $|\dot{\nu}| < 9 \times 10^{-9} \text{ Hz s}^{-1}$ within our O1 observation time spanning 116 days.
The Galactic Center is at a lower declination with respect to the orbital plane of the Earth. The Earth modulation term is therefore more significant so for the Galactic Center we choose combined bins of 0.53 Hz across the band. In this case we are sensitive to a frequency modulation in the range $|\dot{\nu}| < 5.3\times 10^{-8} \text{ Hz s}^{-1}$.

{\em Significance.}---%
To assess the significance of the SNR in the combined bins of the narrowband radiometer spectra, we simulate many realizations of the strain power consistent with Gaussian noise in each individual frequency bin. 
Combining these in the same way as the actual analysis leaves us with a distribution of maximum SNR values across the whole frequency band for many simulations of noise.

For a map of the whole sky, the distribution of maximum \ac{SNR} is complicated by the many dependent trials due to covariances between different pixels (or patches) on the sky. 
We calculate this distribution numerically by simulating many realizations of the dirty map $X_\nu$ with expected covariances described by the Fisher matrix $\Gamma_{\mu\nu}$ (cf.~Eqs.~\ref{eq:Xnu} and ~\ref{eq:Fisher}, respectively). 
This distribution is then used to calculate the significance (or p-value) of a given \ac{SNR} recovered from the sky maps \cite{sph_results}. 
We take a p-value of 0.01 or less to indicate a significant result. 
The absence of any significant events indicates the data are consistent with no signal being detected, in which case we quote Bayesian upper limits at 90\% confidence \cite{radiometer, sph_results}

\begin{figure*}[htbp!]
  \begin{tabular}{ccc}
  	\includegraphics[width=2.15in]{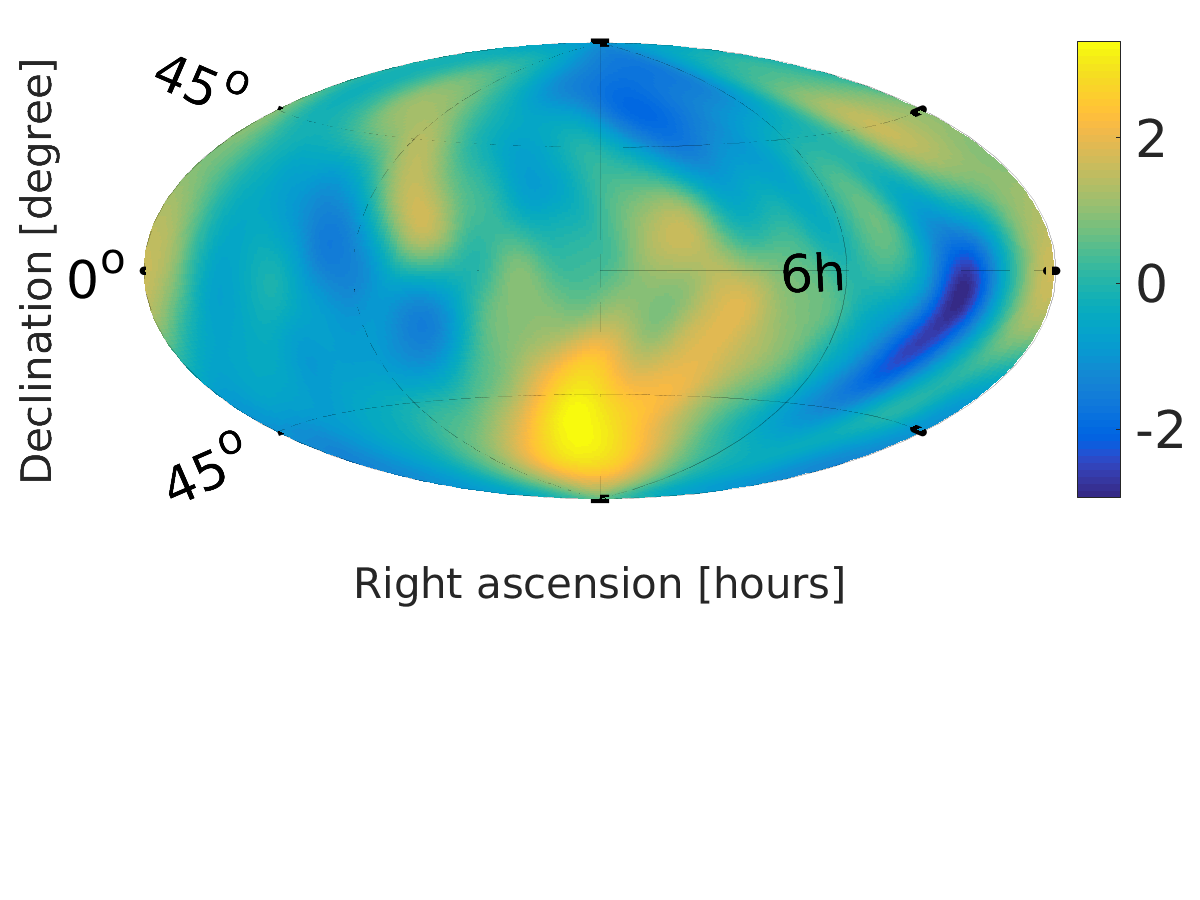} \hspace{.4cm} &
  	\includegraphics[width=2.15in]{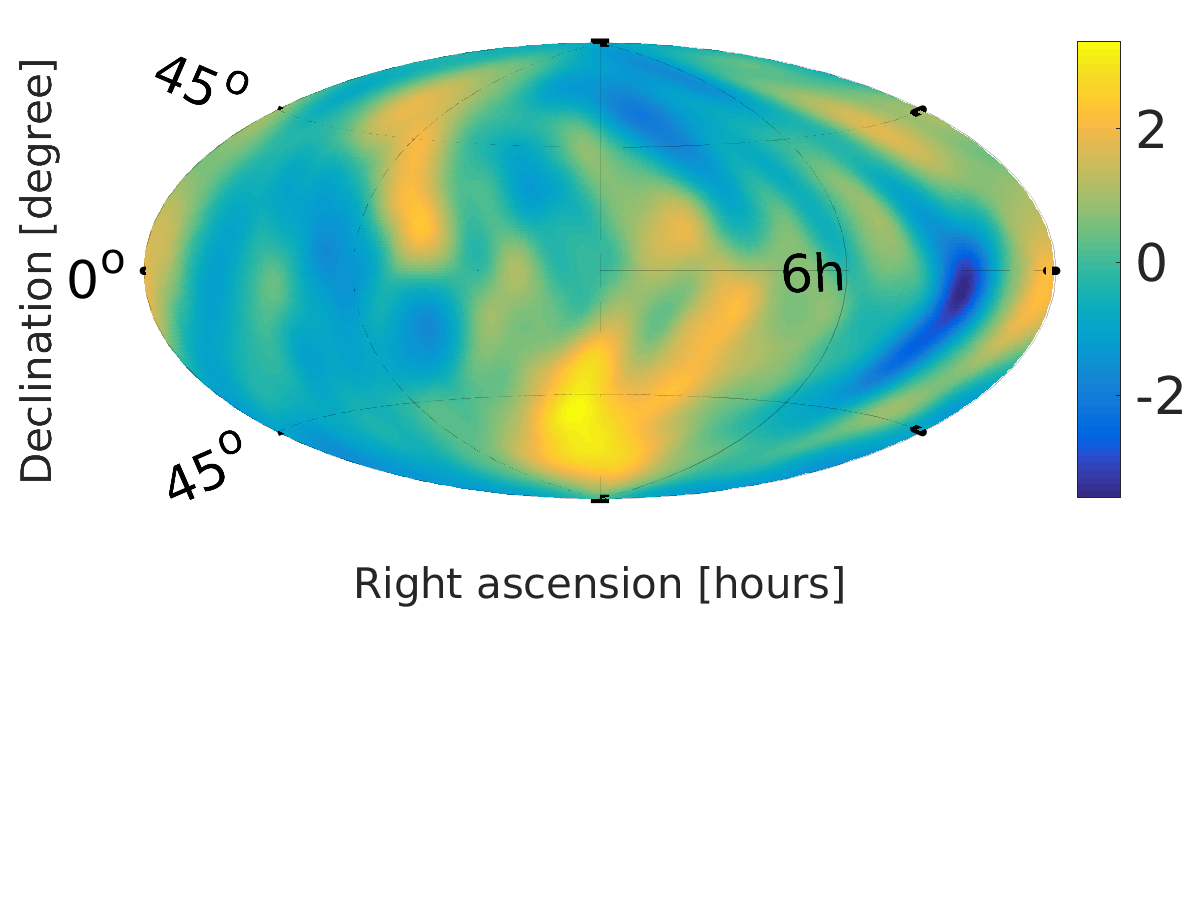} \hspace{.4cm} &
  	\includegraphics[width=2.15in]{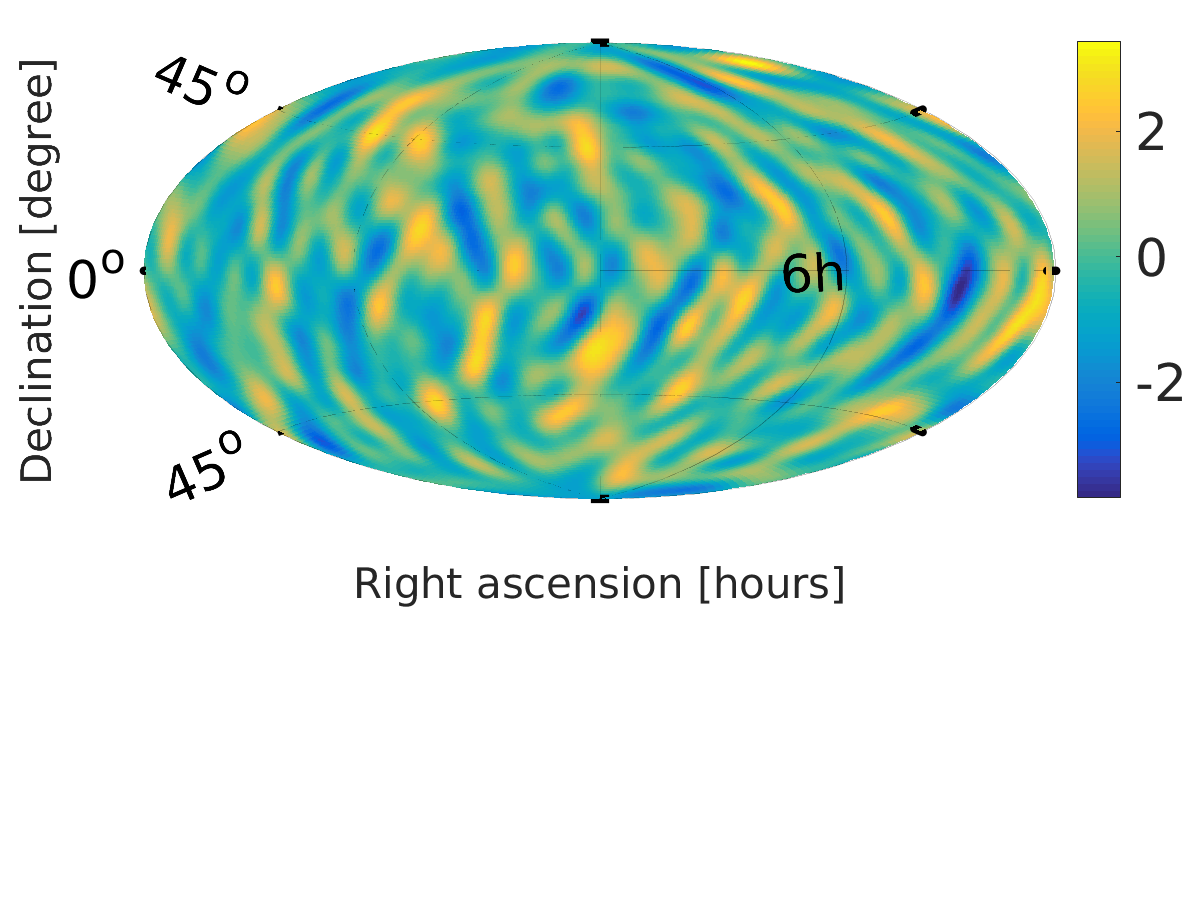} \hspace{.4cm} \vspace {-1.8cm}\\
  	\includegraphics[width=2.3in]{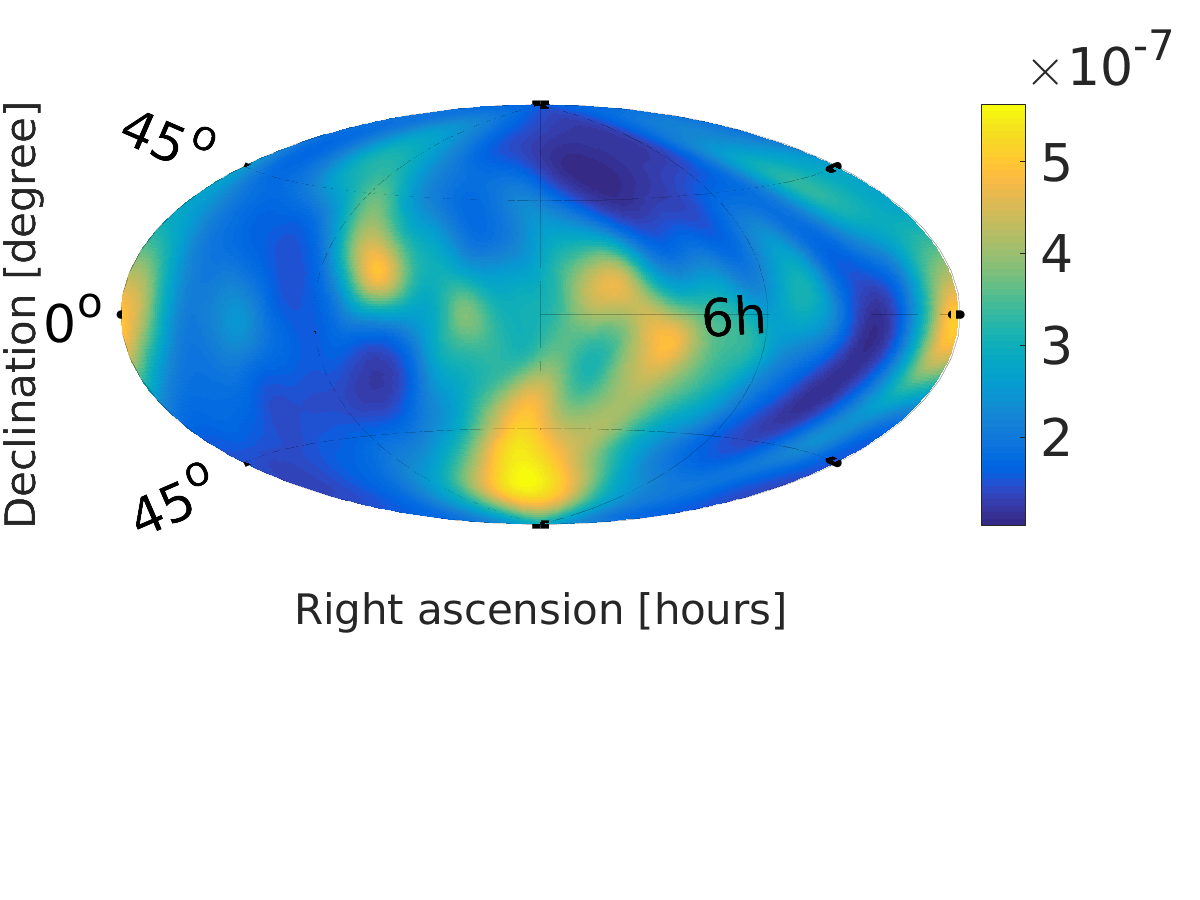} &
  	\includegraphics[width=2.3in]{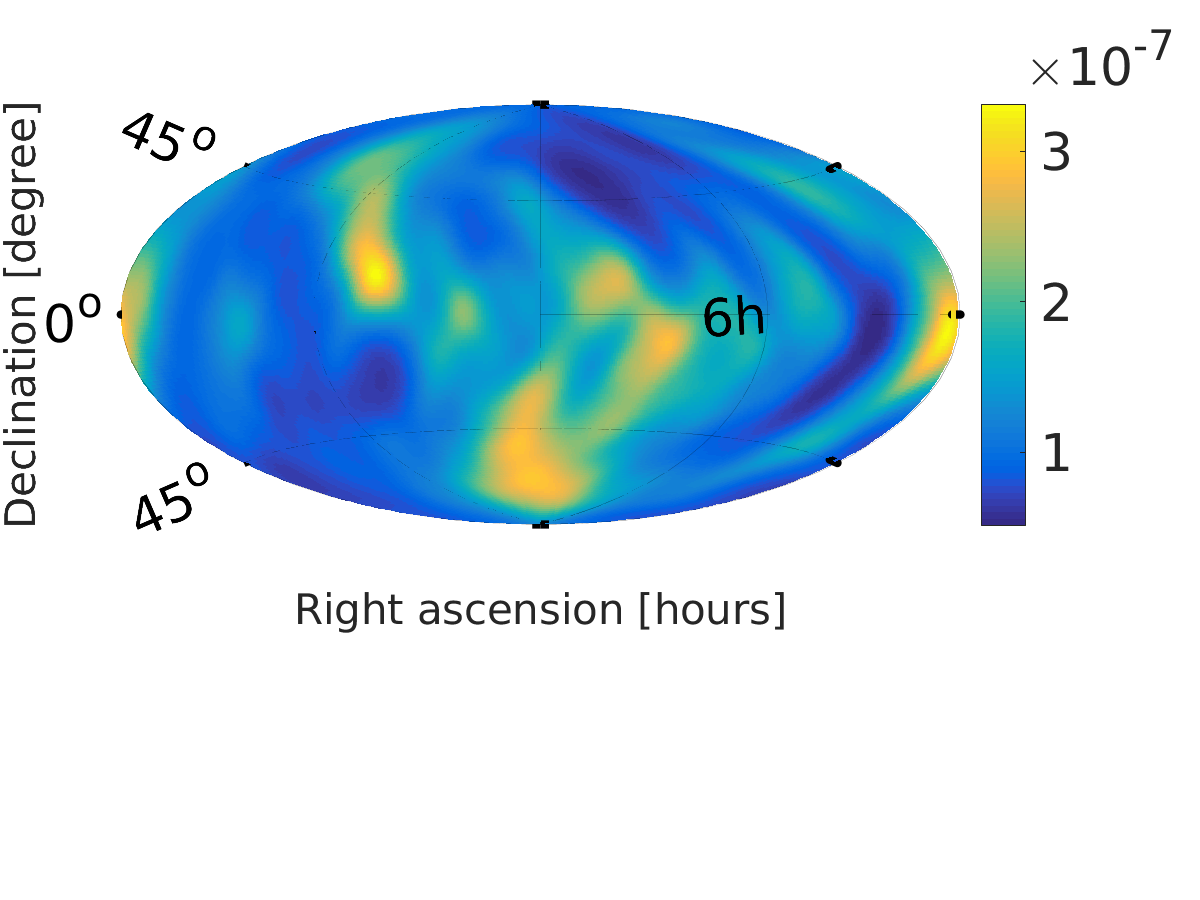} &
  	\includegraphics[width=2.3in]{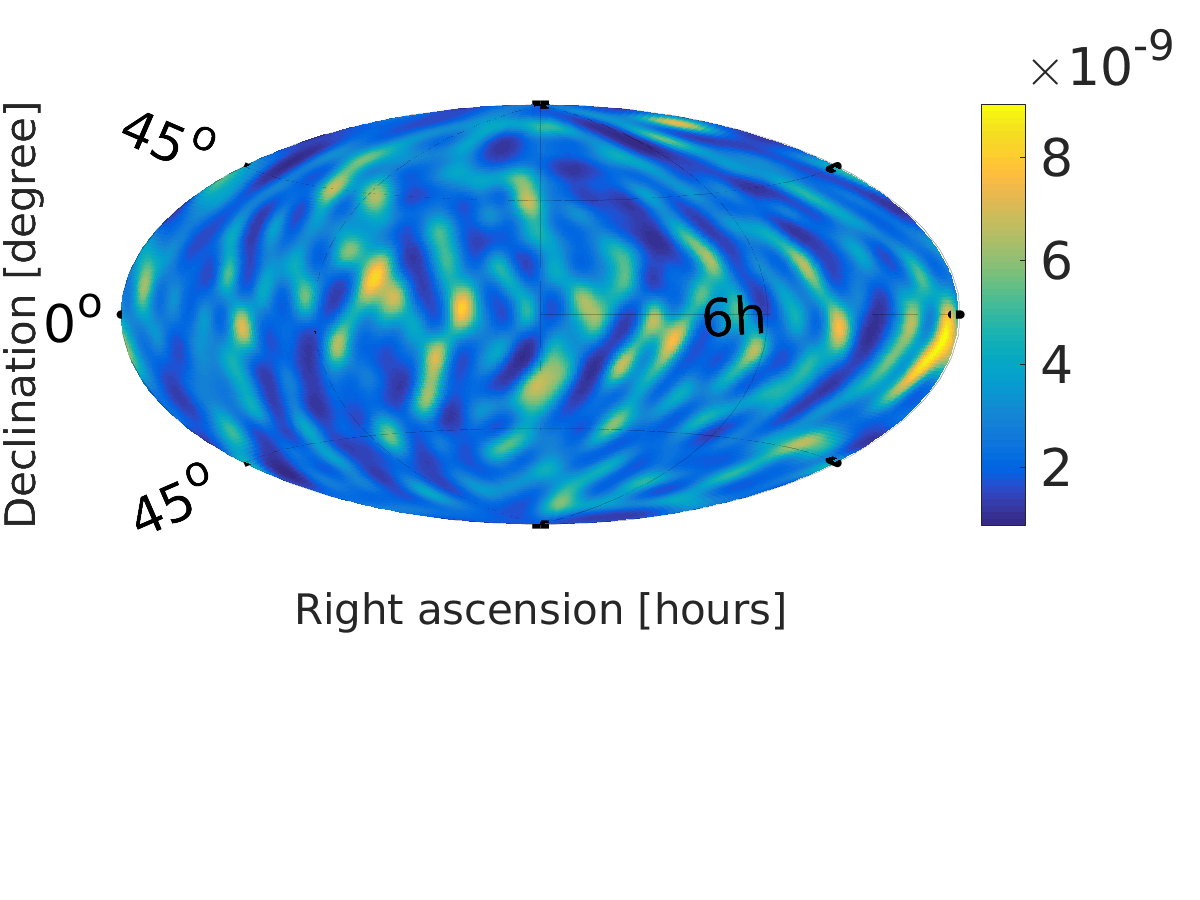} \vspace {-1.6cm}\\
  \end{tabular}
  \caption{
All-sky radiometer maps for point-like sources showing SNR (top) and upper limits at 90\% confidence on energy flux $F_{\alpha,\Theta_0}$ [$\text{erg cm}^{-2}\text{ s}^{-1} \text{ Hz}^{-1}$] (bottom) for three different power-law indices, $\alpha = 0, \, 2/3$ and 3, from left to right, respectively.
The p-values associated with the maximum \ac{SNR} are (from left to right) $p=7\%,\,p=12\%,\,p=47 \%$ (see Table \ref{tab:alphas}). }  
\label{fig:RADmaps}  
\end{figure*}

{\em Results}---The search yields four data products:

{\bf Radiometer sky maps}, optimized for broadband point sources, are shown in Fig.~\ref{fig:RADmaps}.
The top row shows the \ac{SNR}.
Each column corresponds to a different spectral index, $\alpha = 0,\, 2/3$ and 3, from left to right, respectively.
The maximum \acp{SNR} are respectively $3.32$, $3.31$, and $3.43$ corresponding to false-alarm probabilities typical of what would be expected from Gaussian noise; see Table~\ref{tab:alphas}. 
We find no evidence of a signal, and so set limits on gravitational-wave energy flux, which are provided in the bottom row of Fig.~\ref{fig:RADmaps} and summarized in Table~\ref{tab:alphas}.

\begin{figure*}[htbp!]
  \begin{tabular}{ccc}
	\includegraphics[width=2.15in]{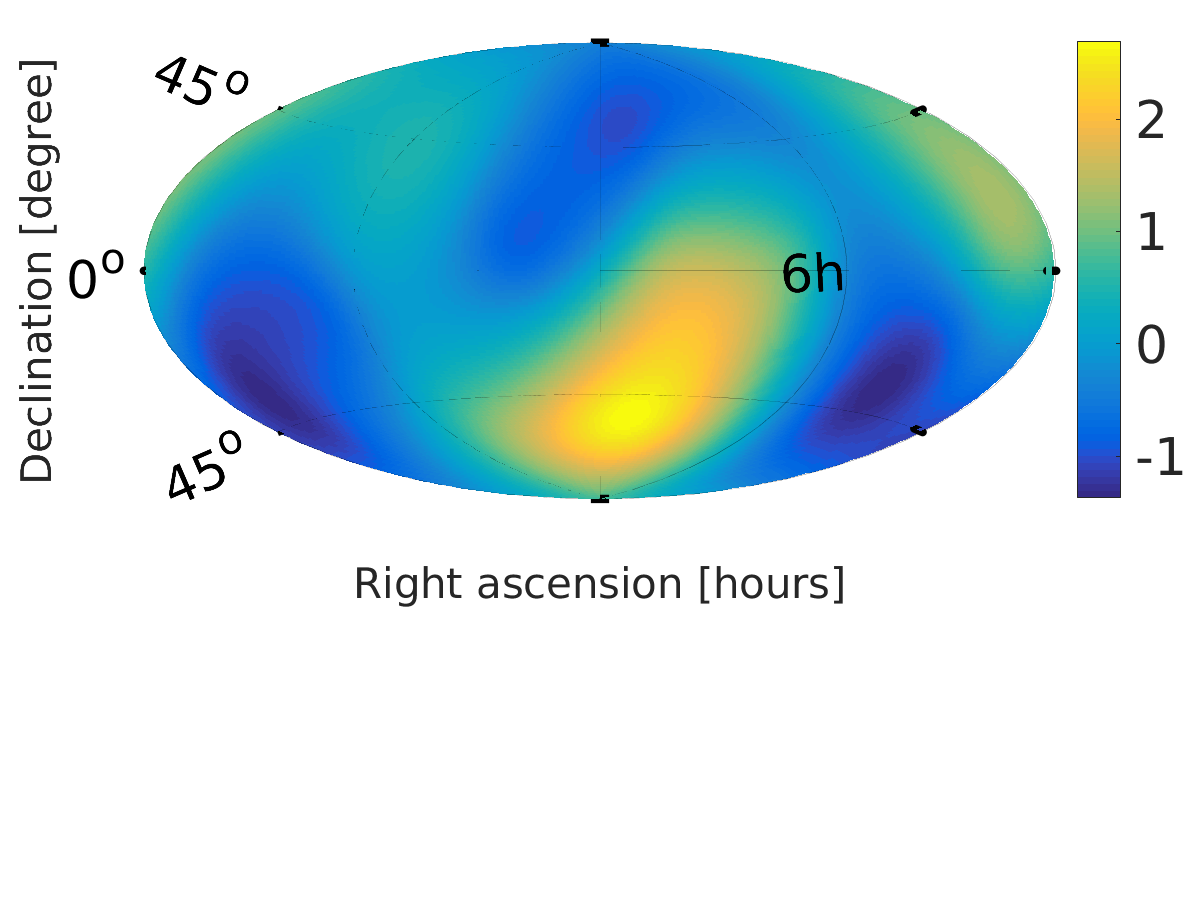} \hspace{.4cm} & 
	\includegraphics[width=2.15in]{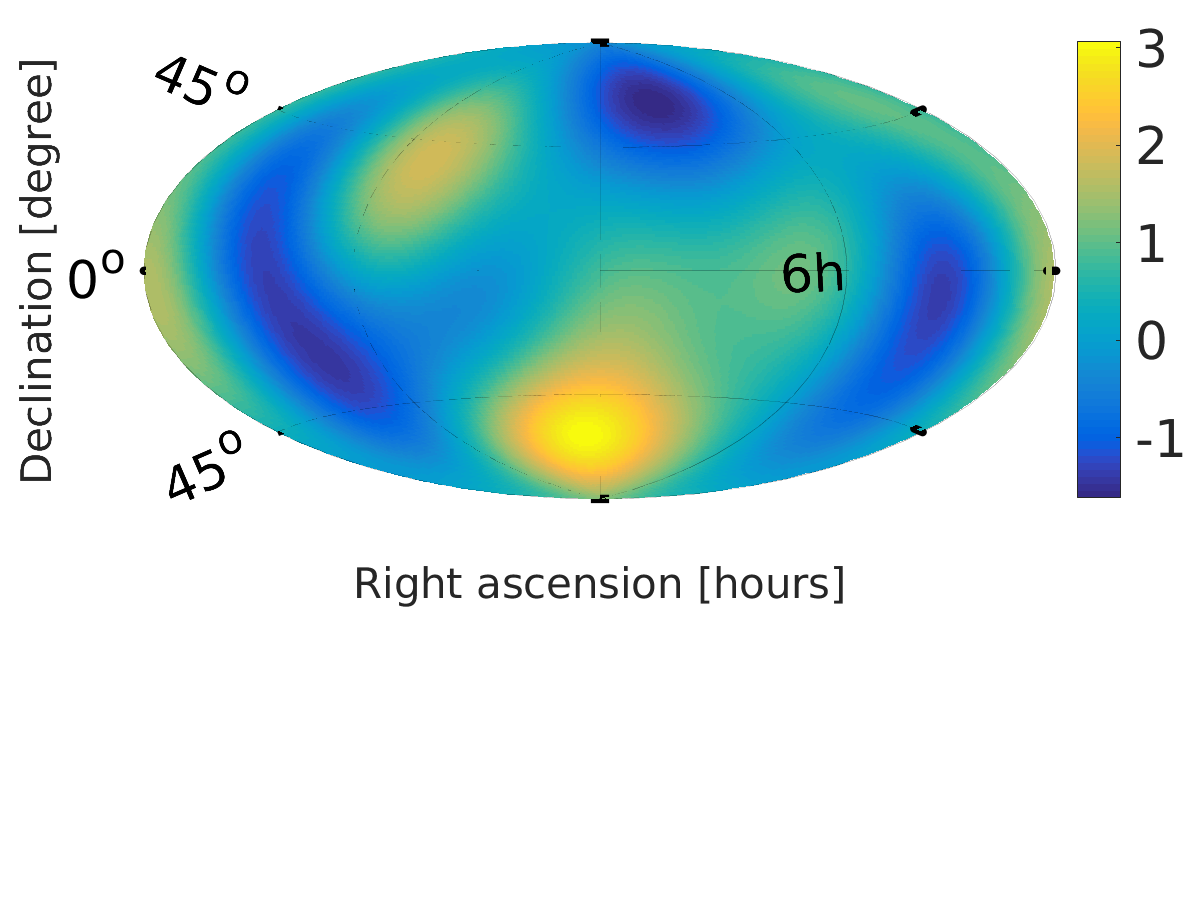} \hspace{.4cm} & 
	\includegraphics[width=2.15in]{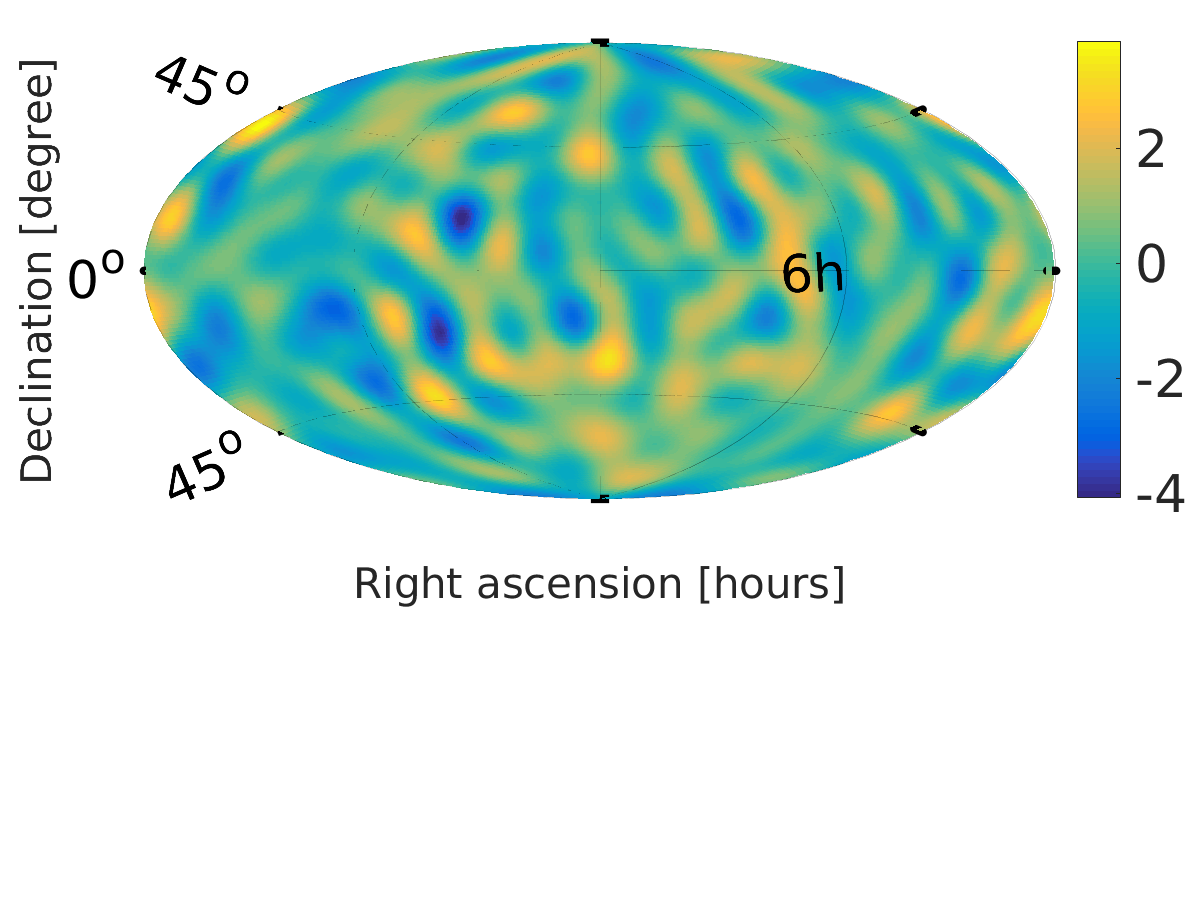} \hspace{.4cm} \vspace {-1.8cm}\\ 
	\includegraphics[width=2.3in]{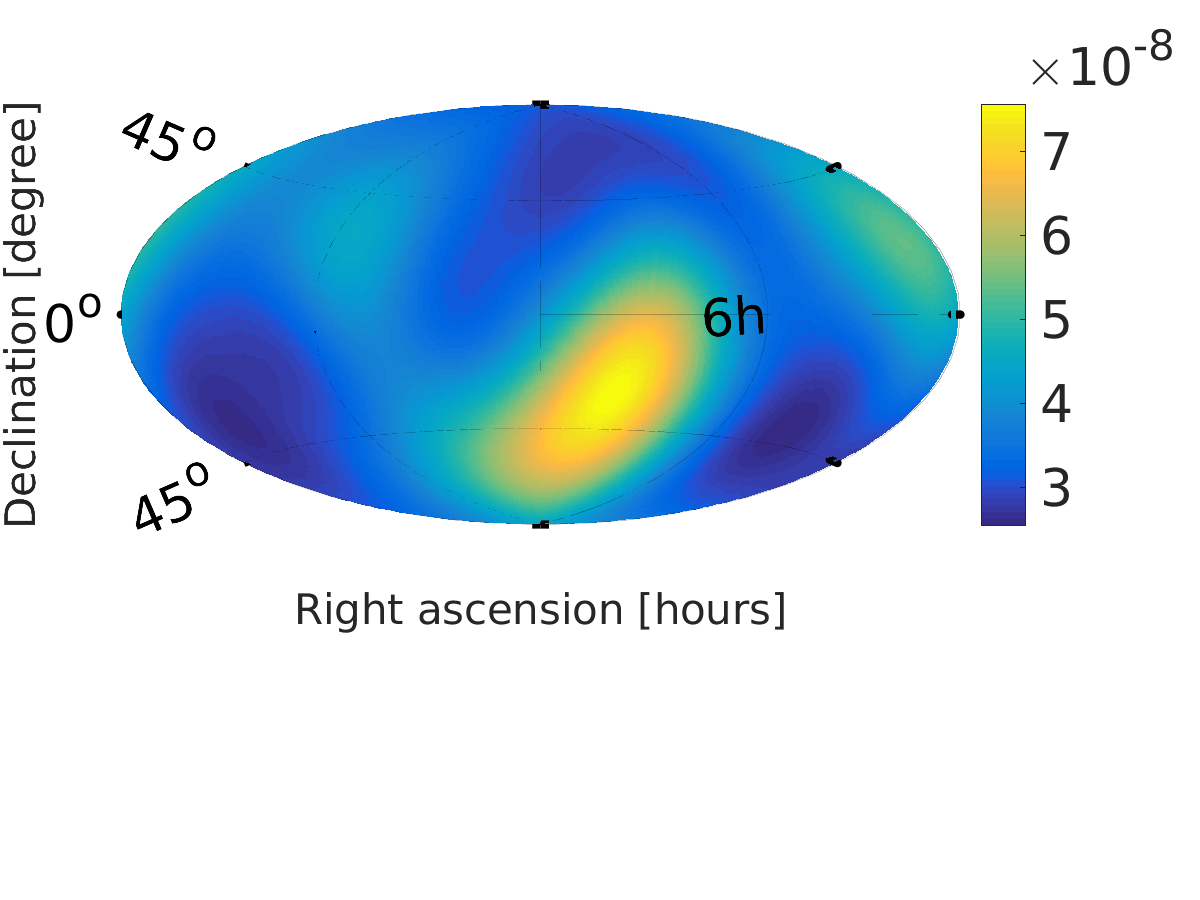} &  
	\includegraphics[width=2.3in]{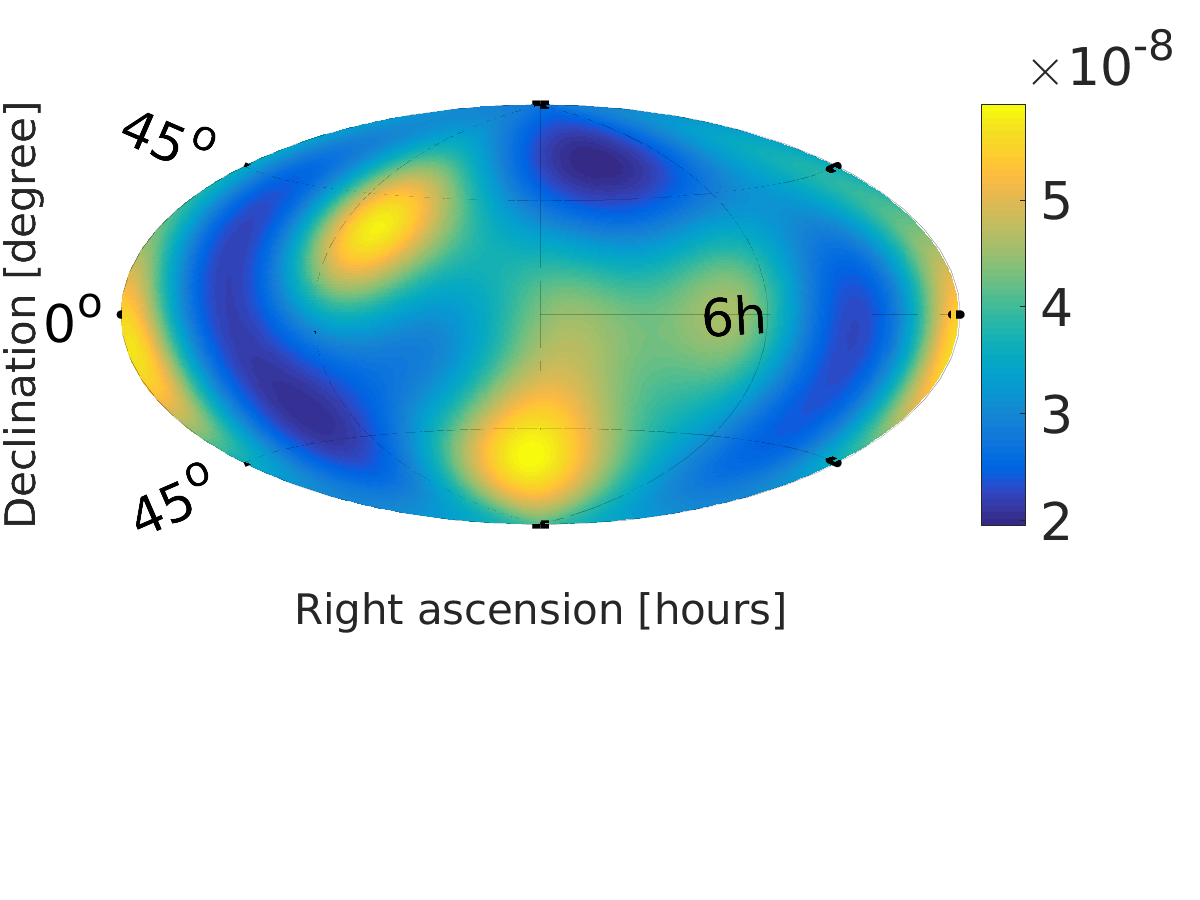} & 
	\includegraphics[width=2.3in]{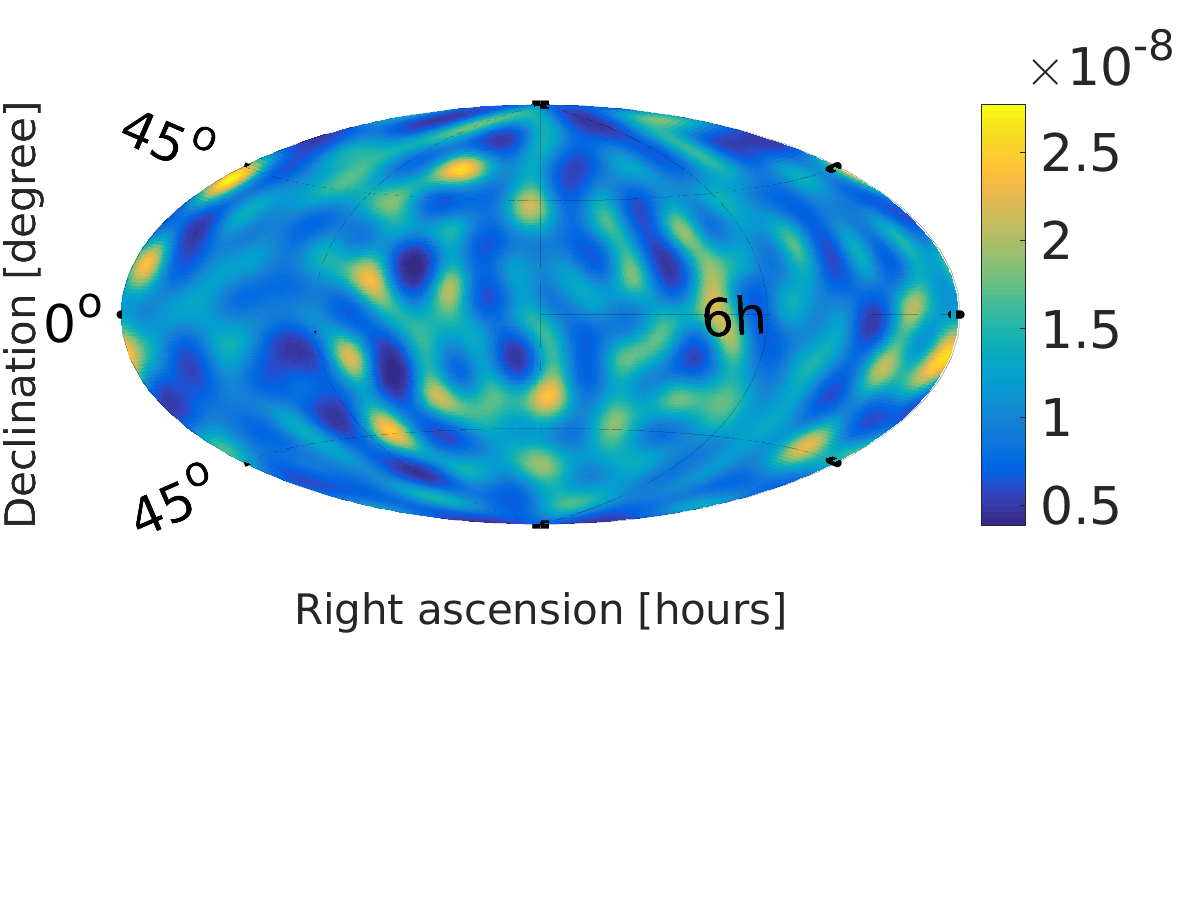} \vspace {-1.6cm} 
  \end{tabular}
  \caption{
All-sky spherical harmonic decomposition maps for extended sources showing SNR (top) and upper limits at 90 \% confidence on the energy density of the gravitational wave background $\Omega_{\alpha}\; [\text{ sr}^{-1}]$ (bottom) for three different power-law indices $\alpha = 0, \, 2/3$ and 3, from left to right, respectively. 
The p-values associated with the maximum \ac{SNR} are (from left to right) $p=18 \%,\,p=11\%,\,p=11\%$ (see Table \ref{tab:alphas}).  }  
\label{fig:SPHmaps}  
\end{figure*}

{\bf SHD sky maps}, suitable for characterizing an anisotropic stochastic background, are shown in Fig.~\ref{fig:SPHmaps}.
The top row shows the \ac{SNR} and each column corresponds to a different spectral index ($\alpha = 0,\,2/3$ and 3, respectively). 
The maximum \acp{SNR} are $2.96$, $3.06$, and $3.86$ corresponding to false-alarm probabilities typical of those expected from Gaussian noise; see Table~\ref{tab:alphas}.
Failing evidence of a signal, we set limits on energy density per unit solid angle, which are provided in the bottom row of Fig.~\ref{fig:SPHmaps} and summarised in Table~\ref{tab:alphas}. 
Interactive visualizations of the SNR and upper limit maps are also available online \cite{skymaps}.
 
\begin{figure}[h!]
   \includegraphics[width=3.5in]{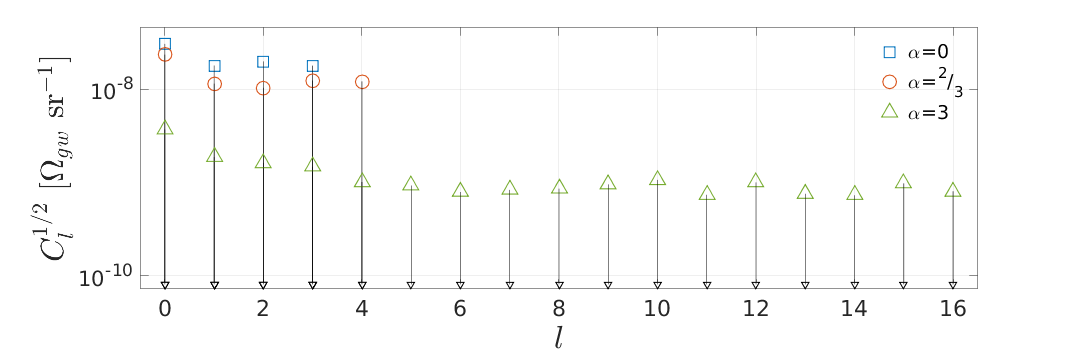}
  \caption{
    Upper limits on $C_l$ at 90\% confidence vs $l$ 
    for the SHD analyses for $\alpha=0$ (top, blue squares), $\alpha=2/3$ (middle, red circles) and $\alpha=3$ (bottom, green triangles).
    \label{fig:Cls} }
\end{figure}

{\bf Angular power spectra} are derived from the SHD sky maps.
We present upper limits at 90\% confidence on the angular power spectrum indices $C_l$ from the spherical harmonic analysis in Figure~\ref{fig:Cls}.

\begin{table*}
  \begin{tabular}{ccc|ccc|cc}
   \multicolumn{3}{c}{} &\multicolumn{5}{c}{\bf Narrowband Radiometer Results} \\
   \multicolumn{2}{c}{Direction}&& Max SNR & p-value (\%) & Frequency band (Hz) & Best UL ($\times 10^{-25}$) & Frequency band (Hz) \\
    \hline 
    &Sco X-1    		&& 4.58  & 10 & $616-617$   & 6.7 & $134-135$ \\
    &SN1987A			&& 4.07  & 63 & $195-196$   & 5.5 & $172-173$ \\
    &Galactic Center   	&& 3.92  & 87 & $1347-1348$ & 7.0 & $172-173$ \\ \hline
  \end{tabular}
  \caption{Results for the narrowband radiometer showing the maximum SNR, corresponding p-value and 1 Hz frequency band as well as the 90\% gravitational wave strain upper limits, and corresponding frequency band, for three sky locations of interest. The best upper limits are taken as the median of the most sensitive 1 Hz band.} \label{tab:rad_nb}
\end{table*}

\begin{figure*}
  \begin{tabular}{ccc}
  \includegraphics[width=2.5in]{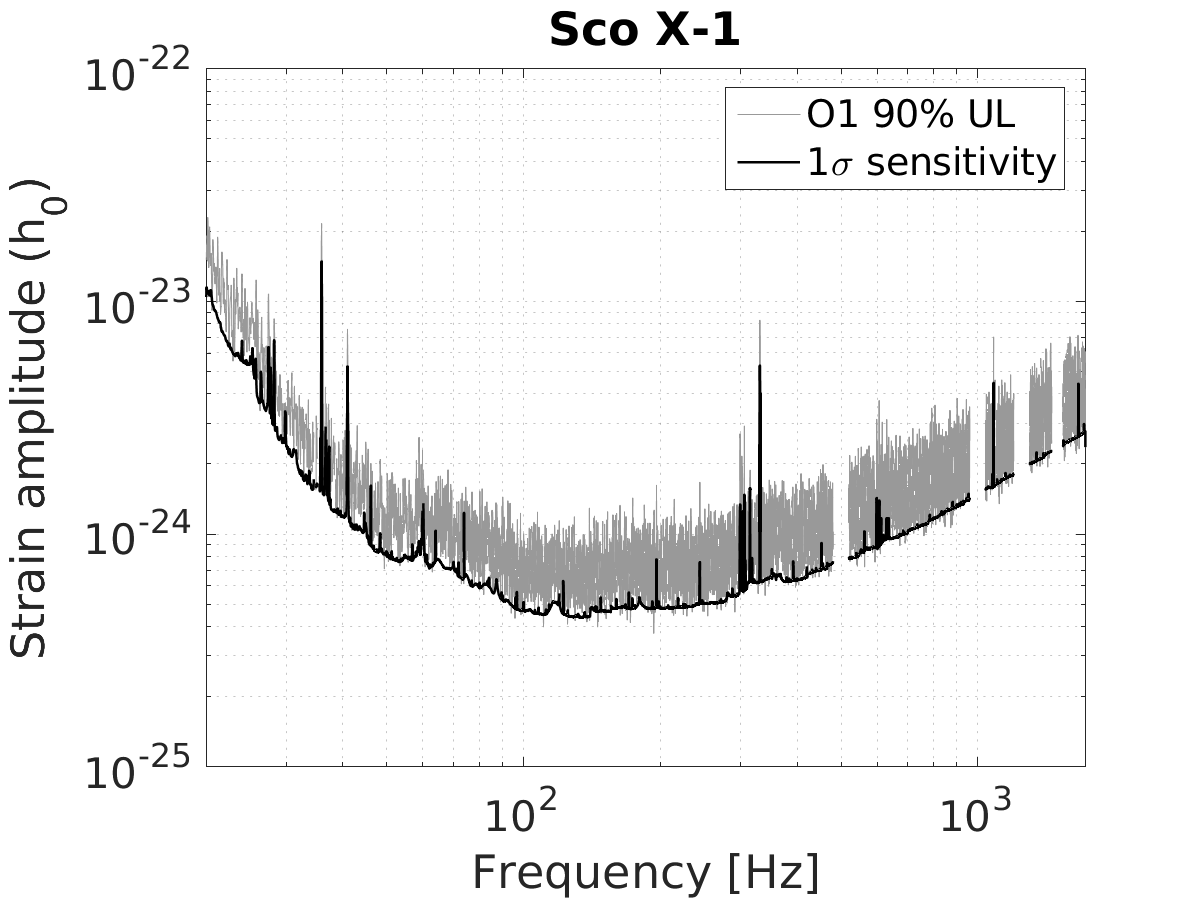} & \hspace{-.4cm}
  \includegraphics[width=2.5in]{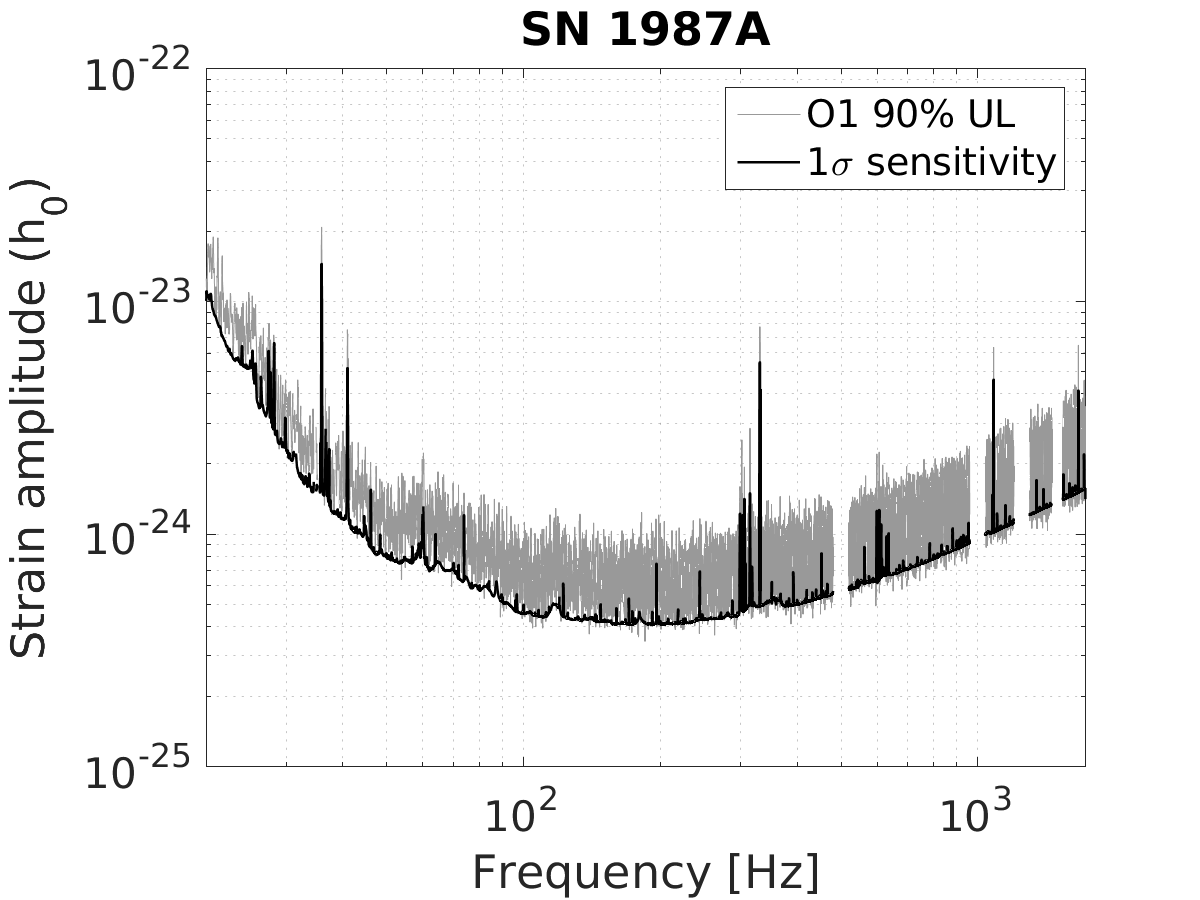} & \hspace{-.4cm}
  \includegraphics[width=2.5in]{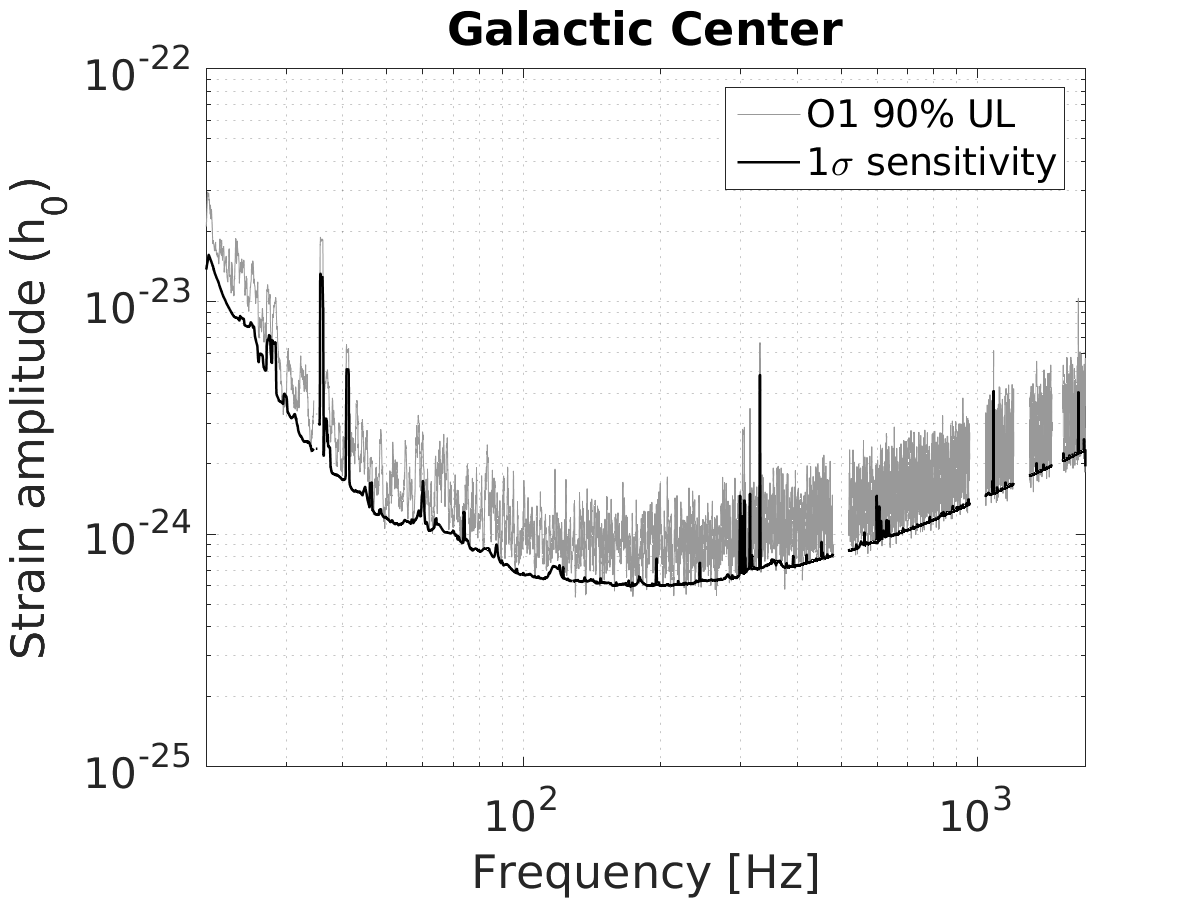} \\
  \end{tabular}
  \caption{
Radiometer 90\% upper limits on dimensionless strain amplitude ($h_0$) as a function of frequency for Sco X-1 (left), SN1987A (middle) and the Galactic Center (right) for the O1 observing run (gray band) and standard deviation $\sigma$ (black line).  
The large spikes correspond to harmonics of the 60~Hz power mains, calibration lines and suspension-wire resonances.
 } \label{fig:rad_directional}
\end{figure*}

{\bf Radiometer spectra}, suitable for the detection of a narrowband point source associated with a given sky position, are given in Fig.~\ref{fig:rad_directional}, the main results of which are summarised in Table \ref{tab:rad_nb}.
For the three sky locations (Sco X-1, SN 1987A and the Galactic Center), we calculate the \ac{SNR} in appropriately sized combined bins across the LIGO band.
For Sco X-1, the loudest observed SNR is 4.58, which is consistent with Gaussian noise. 
For SN 1987A and the Galactic Center, we observe maximum \acp{SNR} of 4.07 and 3.92 respectively, corresponding to false alarm probabilities consistent with noise; see Table \ref{tab:rad_nb}.

Since we observe no statistically significant signal, we set 90\% confidence limits on the peak strain amplitude $h_0$ for each optimally sized frequency bin. 
Upper limits were set using a Bayesian methodology with the constraint that $h_0 > 0$ and validated with software injection studies. 
The upper limit procedure is described in more detail in the technical supplement \cite{supplement_nb}, while the subsequent software injection validation is detailed in \cite{techdoc_nb}.

The results of the narrowband radiometer search for the three sky locations are shown in Fig.~\ref{fig:rad_directional}. 
To avoid setting limits associated with downward noise fluctuations, we take the median upper limit from the most sensitive 1 Hz band as our best strain upper limit.
We obtain $90\%$ confidence upper limits on the gravitational wave strain of \radULsco, \radULgc and \radULsn from Sco X-1, SN 1987A and the Galactic Center respectively in the most sensitive part of the LIGO band 

{\em Conclusions.} 
We find no evidence to support the detection of either point-like or extended sources and set upper limits on the energy flux and energy density of the anisotropic gravitational wave sky. 
We assume three different power law models for the gravitational wave background spectrum. 
Our mean upper limits present an improvement over initial LIGO results of a factor of {$8$} in flux for the $\alpha=3$ broadband radiometer and factors of {$60$} and {$4$} for the spherical harmonic decomposition method for $\alpha=0$ and 3 respectively \cite{sph_results,compareS5}. 
We present the first upper limits for an anisotropic stochastic background dominated by compact binary inspirals (with an $\Ogw \propto f^{2/3}$ spectrum) of $\Omega_{2/3}(\ohat) < 2 - 6 \times 10^{-8} \text{ sr}^{-1}$ depending on sky position. 
We can directly compare the monopole moment of the spherical harmonic decomposition to the isotropic search point estimate $\Omega_{2/3} = (3.5 \pm 4.4) \times 10^{-8}$ from \cite{o1iso}. 
We obtain $\Omega_{2/3} = ({2\pi^2}/{3H_0^2})\fref^3 \sqrt{4\pi} {\cal P}_{00} = (4.4 \pm 6.4) \times 10^{-8}$.
The two results are statistically consistent. Our spherical harmonic estimate of $\Omega_{2/3}$ has a larger uncertainty than the dedicated isotropic search because of the larger number of (covariant) parameters estimated when $l_\text{max}>0$.
We also set upper limits on the gravitational wave strain from point sources located in the directions of Sco X-1, the Galactic Center and Supernova 1987A. 
The narrowband results improve on previous limits of the same kind by more than a factor of {$10$} in strain at frequencies below 50 Hz and above 300 Hz, with a mean improvement of a factor of {2} across the band \cite{sph_results}.

\emph{Acknowledgments.}---{\small The authors gratefully acknowledge the support of the United States
National Science Foundation (NSF) for the construction and operation of the
LIGO Laboratory and Advanced LIGO as well as the Science and Technology Facilities Council (STFC) of the
United Kingdom, the Max-Planck-Society (MPS), and the State of
Niedersachsen/Germany for support of the construction of Advanced LIGO 
and construction and operation of the GEO600 detector. 
Additional support for Advanced LIGO was provided by the Australian Research Council.
The authors gratefully acknowledge the Italian Istituto Nazionale di Fisica Nucleare (INFN),  
the French Centre National de la Recherche Scientifique (CNRS) and
the Foundation for Fundamental Research on Matter supported by the Netherlands Organisation for Scientific Research, 
for the construction and operation of the Virgo detector
and the creation and support  of the EGO consortium. 
The authors also gratefully acknowledge research support from these agencies as well as by 
the Council of Scientific and Industrial Research of India, 
Department of Science and Technology, India,
Science \& Engineering Research Board (SERB), India,
Ministry of Human Resource Development, India,
the Spanish Ministerio de Econom\'ia y Competitividad,
the Conselleria d'Economia i Competitivitat and Conselleria d'Educaci\'o, Cultura i Universitats of the Govern de les Illes Balears,
the National Science Centre of Poland,
the European Commission,
the Royal Society, 
the Scottish Funding Council, 
the Scottish Universities Physics Alliance, 
the Hungarian Scientific Research Fund (OTKA),
the Lyon Institute of Origins (LIO),
the National Research Foundation of Korea,
Industry Canada and the Province of Ontario through the Ministry of Economic Development and Innovation, 
the Natural Science and Engineering Research Council Canada,
Canadian Institute for Advanced Research,
the Brazilian Ministry of Science, Technology, and Innovation,
Funda\c{c}\~ao de Amparo \`a Pesquisa do Estado de S\~ao Paulo (FAPESP),
Russian Foundation for Basic Research,
the Leverhulme Trust, 
the Research Corporation, 
Ministry of Science and Technology (MOST), Taiwan
and
the Kavli Foundation.
The authors gratefully acknowledge the support of the NSF, STFC, MPS, INFN, CNRS and the
State of Niedersachsen/Germany for provision of computational resources.
This is LIGO document \dcc. }

\bibliography{sph_O1results}

\begin{thebibliography}{41}
\bibitem [{com()}]{compareS5}%
  \BibitemOpen
  \href@noop {} {}\bibinfo {note} {The spherical harmonic analysis uses
  different values of the maximum order $\lmax$ than those used in
  \cite{sph_results}. The $\lmax$ value was decreased by a factor of 2 for
  $\alpha=0$ and increased by $30\%$ for $\alpha=3$. The harmonic order
  resembles extra degrees of freedom, meaning that larger values of $\lmax$
  tend to reduce the sensitivity in $\Ogw$ by increasing uncertainty in
  estimates of the angular power $\ophat$.}\BibitemShut {Stop}%
\end{thebibliography}

%
%
%
%
%
%
%
%
%
%

\section*{Supplement--Directional limits on persistent gravitational waves from Advanced LIGO's first observing run}

In this supplement we describe how we use the narrowband, directed radiometer search\cite{radio_method} to make a statement on the gravitational wave (GW) strain amplitude $h_0$ of a persistent source given some power described by our cross correlation statistic. 
We take into account the expected modulation of the quasi-monochromatic source frequency over the duration of the observation. 
We do this by combining individual search frequency bins into \emph{combined bins} that cover the extent of the possible modulation.

{\em Source Model.}---We can relate GW frequency emitted in the source frame $\fs$ to the observed frequency in the detector frame $\fdet$ using the relation
\begin{equation}
\fdet = [1 - A(t) - B(t) - C(t)]\fs \label{eq:detector_to_source}
\end{equation}
where $A(t)$ takes into account the modulation of the signal due to Earth's motion with respect to the source, $B(t)$ takes into account the orbital modulation for a source in a binary orbit, and $C(t)$ takes into account any other modulation due to intrinsic properties of the source (for example any spin-down terms for isolated neutron stars).

The Earth modulation term is given by 
\begin{equation}
A(t) = \frac{\ve(t)\cdot{\hat{k}}}{c} 
\end{equation}
where $\ve$ is the velocity of the Earth. In equatorial coordinates:
\begin{equation}
\ve(t) = \omega R\left[\sin\theta(t)\hat{u} - \cos\theta(t)\cos\phi\hat{v} - \cos\theta(t)\sin\phi\hat{w}\right], \\
\end{equation}
in which $R$ is the mean distance between Earth and the Sun, $\omega$ is the angular velocity of the Earth around the Sun and $\phi=$ 23$^\circ$, 26 min, 21.406 sec is the obliquity of the ecliptic. The time dependent phase angle $\theta(t)$  is given by $\theta(t)=2\pi(t-T_\text{VE})/T_\text{year}$,
where $T_\text{year}$ is the number of seconds in a year and $T_\text{VE}$ is the time at the Vernal equinox.  
The unit vector $\hat{k}$ pointing from the source to the earth is given by
$\hat{k} = -\cos\delta \cos\alpha \hat{u} - \cos\delta\sin\alpha\hat{v}-\sin\delta\hat{w}$, where $\delta$ is the declination and $\alpha$ is the right ascension of the source on the sky.

In the case of a source in a binary system, the binary term (for a circular orbit) is given by
\begin{equation}
B(t) = \frac{2\pi}{\Porb} a\sin i \times \cos\left(2\pi\frac{t-\Tasc}{\Porb}\right)
\end{equation}
where $a\sin i$ is the projection of the semi-major axis (in units of light seconds) of the binary orbit on the line of sight, $\Tasc$ is the time of the orbital ascending node and $\Porb$ is the binary orbital period.

In the case of an isolated source we set $B(t)=0$, while $C(t)$ can take into account any spin modulation expected to occur during an observation time. In the absence of a model for this behaviour, a statement can be made on the maximum allowable spin modulation that can be tolerated by our search.



{\em Search.}---The narrowband radiometer search is run with 192 s segments and 1/32 Hz frequency bins. 
For each 1/32 Hz {frequency bin} we combine the number of bins required to account for the extent of any signal frequency modulation
The source frequency $\fs$ is taken as the center of a {frequency bin}. 
We calculate the minimum and maximum detector frequency $\fd$ over the time of the analysis corresponding to the respective \emph{edges} of the bin in order to define our {combined bins}.

We combine the detection statistic  $Y_i$ and variance $\sigma_{Y,i}$ into a new combined statistic $Y_c$ for each representative frequency bin via
%
%
\begin{equation}
        Y_c = \sum_{i=-b}^{a} Y_i \;\;  \text{and} \;\;  \sigma^2_{Y,c} = \sum_{i=-b}^{a} \sigma^2_{Y,i} \, ,\label{eq:Yc}
\end{equation}
where $i$ represents the index for each of the {frequency} bins we want to combine. If we assign $i=0$ to the bin where the source frequency falls, then $a$ and $b$ are the number of frequency bins we want to combine above and below the source frequency bin, respectively. 
The overlapping bins, which ensure we do not lose signal due to edge effects, create correlations between our \emph{combined} bins.

{\em Significance.}---To establish significance, we assume that the strain power in each {frequency bin} is consistent with Gaussian noise and simulate $>1000$ noise realizations. For each realization, we generate values of $Y_i$ in each {frequency bin} $i$ by drawing from a Gaussian distribution with $\sigma = \sigma_{Y,i}$. We then combine these bins into {combined bins} as we do in the actual analysis and calculate the maximum of the signal to noise ratio, $\text{SNR} = Y_c / \sigma_{Y,c}$, across all of the {combined bins}.
We use the distribution of maximum SNR to establish the significance of our results.


{\em Upper limits.}---In the absence of a significant detection statistic, we set upper limits on the tensor strain amplitude $h_0$ of a gravitational wave source with frequency $\fs$. 
To take into account the unknown parameters of the system, such as the polarization $\psi$ and inclination angle $\iota$, and consider reduced sensitivity to signals that are not circularly polarized, we calculate a direction-dependent and time-averaged value $\mu_{\iota,\psi}$. 
This value 
represents a scaling between the true value of the amplitude $h_0$ and what we would measure with our search, 
and is given by
\begin{equation}
\mu_{\iota,\psi} = \frac{\sum_{j=1}^M \left[(A^+/h_0)^2 F^+_{\text{d}j} + (A^{\times}/h_0)^2 F^{\times}_{\text{d}j} \right]\left(F^+_{\text{d}j} + F^\times_{\text{d}j} \right)}{\sum_{j=1}^M \left(F^+_{\text{d}j} + F^\times_{\text{d}j} \right)^2}
\end{equation}
for each time segment $j$. Here 
\begin{equation}
        A^+ = \frac{1}{2}h_0(1+\cos^2\iota) \;\;  \text{and} \;\;  A^\times = h_0 \cos\iota \, ,\label{eq:Apc}
\end{equation} 
and $\psi$ dependence is implicit in $F^A_{\text{d}j} = F^A_{1j} F^A_{2j}$, where $A$ indicates ($+$ or $\times$) polarization and the response functions $F^A_{1j}$ and $F^A_{2j}$ for the LIGO detectors are defined in \cite{allen-romano} (see also \cite{chris}). %
We calculate $\mu_{\iota,\psi}$ many times for a uniform distribution of $\cos\iota$ and $\psi$, and then marginalize over it. We also marginalize over calibration uncertainty, where we assume (as in the past) that calibration uncertainty is manifest in a multiplicative factor $(l + 1) > 0$ where $l$ is normally distributed around 0 with uncertainty given by the calibration uncertainty, $\sigma_l = 0.18$. The full expression for our posterior distribution given a measurement, $Y$ and its uncertainty $\sigma_Y$ in a single {combined bin} is given by
\begin{equation}
        p(h_0|Y,\sigma_Y) = \int_{-1}^{1}d(\cos\iota) \int_{-\pi/4}^{\pi/4}d\psi\,\int_{-1}^\infty dl\,e^{L(l)},
\end{equation}
where 
\begin{equation}
L(l) = -\frac{1}{2} \left\{ \left(\frac{l}{\sigma_l}\right)^2 + \left[\frac{Y(l+1) - \mu_{\iota,\psi}h_0^2}{\sigma_Y(l+1)}\right]^2 \right\}.
\end{equation}

We set upper limits on $h_0$ at a $90\%$ credible level for frequencies within each {combined bin} that correspond to each source frequency $\fs$. Denoted $h_0^\text{UL}$, upper limits are calculated via
$0.9  = \int_0^{h_0^\text{UL}} dh_0\,  p(h_0|Y,\sigma_Y).$

{\em Frequency Notches.}---For frequency bins flagged to be removed from the analysis due to instrumental artifacts, we set our statistic to zero, so they will not contribute to the combined statistics described in Eq \ref{eq:Yc}.
We require that more than half of the frequency bins are still available when generating a combined bin.
When setting upper limits, the noise $\sigma_{Y,c}$ in any {combined bin} that contained notched frequency bins is rescaled to account for the missing bins to provide a more accurate representation of the true sensitivity in that {combined bin}.



\emph{}

\end{document}